\documentclass{aa}
\usepackage{supertabular}
\def\lae{\mathrel{<\kern-1.0em\lower0.9ex\hbox{$\sim$}}}
\def\gae{\mathrel{>\kern-1.0em\lower0.9ex\hbox{$\sim$}}}
\def\ecs{{\rm erg} \, {\rm cm}^{-2} \, {\rm sec}^{-1}} 
\def\grad{ ^{\circ} }
\def\persec{\,{\rm sec}^{-1} }
\def\keV{\,{\rm keV} }
\def\mag{\,{\rm mag}}
\def\whz{\,{\rm W/Hz}}
\def\mJy{\,{\rm mJy}}

\def\Hz{\,{\rm Hz}}
\def\GHz{\,{\rm GHz}}
\def\aox{\alpha_{\rm OX}}
\def\aro{\alpha_{RO}}
\def\arx{\alpha_{\rm RX}}

\usepackage{epsfig}
\begin{document}
\title{The HRX-BL Lac sample - evolution of BL Lac objects}
\author{V. Beckmann\inst{1,}\inst{2,}\inst{3},
  D. Engels \inst{1},  N. Bade \inst{1}, and O. Wucknitz\inst{4}}
\offprints{Volker.Beckmann@obs.unige.ch}
\institute{Hamburger Sternwarte, Gojenbergsweg 112, D-21029 Hamburg, Germany
 \and INTEGRAL Science Data Centre, Chemin d' \'Ecogia 16, CH-1290
 Versoix, Switzerland
 \and Institut f\"ur Astronomie und Astrophysik, Universit\"at
 T\"ubingen, Sand 1, D-72076 T\"ubingen, Germany
 \and Universit\"at Potsdam, Astroteilchenphysik, Am Neuen Palais 10,
 D-14469 Potsdam, Germany}
\date{Received date; accepted date}
\authorrunning{Beckmann et al.}
\titlerunning{The HRX-BL Lac sample}

\abstract{ The unification of X-ray and radio selected BL Lacs
has been an outstanding problem in the blazar research in the past
years. Recent investigations have shown that the gap between the two
classes can be filled with intermediate objects and
that apparently all differences can be explained by mutual shifts of the peak frequencies of the synchrotron and inverse Compton component of the emission. 
We study the consequences of this scheme using a new
sample of X-ray selected BL Lac objects comprising 104 objects
with $z<0.9$ and a mean redshift $\bar{z} = 0.34$. 77 BL Lacs, of
which the redshift could be determined for 64 (83\%) objects, form a
complete sample. 
The new data could not confirm our earlier result, drawn from a
subsample, that the negative evolution vanishes below a synchrotron
peak frequency $\log \nu_{\rm peak} = 16.5$.  The complete sample
shows negative evolution at the 2$\sigma$ level ($\langle V_{\rm
e}/V_{\rm a} \rangle = 0.42 \pm 0.04$). 
We conclude that the observed properties
of the HRX BL Lac sample show typical behaviour for X-ray selected BL Lacs. 
They support an evolutionary model, in which
flat-spectrum radio quasars (FSRQ) with high energetic jets evolve
towards low frequency peaked (mostly radio-selected) BL Lac objects
and later on to high frequency peaked (mostly X-ray selected) BL
Lacs.
 }

\maketitle

\section{Introduction}
BL Lac objects are a rare type of Active Galactic Nuclei (AGN), which
are observationally distinguished mainly by the absence of strong
emission lines. They have strong X-ray and radio emission, and they
often show strong variability and optical polarization. Their
observational properties are usually explained by a particular line of
sight toward the galaxy nucleus, which in BL Lacs is thought to be
parallel to a jet emerging from this nucleus.  

Two search strategies are commonly used to find BL~Lacs.
The first is a search for strong X-ray sources with a high ratio of
X-ray to optical flux, yielding X-ray selected BL Lacs (XBL).
The second is to search among flat
spectrum radio sources to find radio selected ones (RBL).
As the radio and X-ray surveys got more and more sensitive, the properties of
both groups started to
overlap, raising the question of how they are
related.  Padovani \& Giommi (\cite{emss}) noticed that the spectral
energy distribution (quantified by $\log \nu L_\nu$)
of radio and X-ray selected BL~Lacs shows peaks at different frequencies,
and suggested that this  
is the basic difference between the two classes of BL~Lacs. They
introduced the notation of {\em high-energy cutoff BL~Lacs} (HBLs) and
{\em low-energy cutoff BL Lacs} (LBLs).  Most, but not all, XBLs are
HBLs, while the group of LBLs is preferentially selected in the radio
region. 

In general, BL~Lacs are considered as part of a larger class of
objects, the blazars, which have similar properties but show emission
lines in addition, and for which this scenario applies as well.
Ghisellini et al. (\cite{ghisellini2}) proposed that the range
of peak frequencies observed is governed primarily by the efficiency
of radiative cooling, and that the other physical parameters strongly depend
on it.
They found an inverse correlation
between the energy of the Lorentz factor of particles emitting at the
peaks of the SED ($\gamma_{\rm peak}$) and the energy density of the
magnetic and radiation field of $\gamma_{\rm peak} \propto
U^{-0.6}$. This correlation was extended later on for low power (high
peaked) BL Lacs by taking into account 
 the finite time for the injection of
particles in the jet (Ghisellini et al. \cite{ghisellini02}). Combined
modelling of the time-dependent electron injection and the self
consistent radiation transport in jets of high peaked blazars lead to
the conclusion that differences in the appearance can be explained by
either self-synchrotron or external Compton dominated processes
(B\"ottcher \& Chiang \cite{ssc/ec}). Other studies focussed on the
importance of shock events in the blazar jets to explain variability
on short timescales (e.g. Bicknell \& Wagner \cite{bicknell}). Based
on these models the most important factor in the appearance of blazars
seems nowadays the energy density of the jet. Maraschi \& Tavecchio
(\cite{diskjet}) showed that this energy density is related to the
accretion rate in the AGN disk and proposed that all blazar types have
similar black hole masses but that the low power blazars exhibit lower
accretion rates.

Unlike all other AGNs, and different from RBLs also, the space density
or the luminosity of XBLs showed an increase with time (e.g. Rector et
al. \cite{emssbllac}). This is called negative evolution. Bade et
al. (\cite{bade}) probed this property with a new ROSAT selected XBL
sample and confirmed the negative evolution only for extreme XBLs,
e.g. HBLs with very high energy cutoffs ($\log \nu_{\rm peak} >
16.5$).  The result for less extreme XBLs, called intermediate-energy
cutoff BL Lacs (IBLs) by Bade et al., was compatible with no
evolution.  This difference in evolutionary behaviour indicated the
presence of a smooth transition between HBLs and LBLs.  However, these
findings were based on only 39 BL Lacs, prompting us to increase the
size of this sample considerably. The results of this effort, the
HRX-BL\,Lac sample presented here, comprises now 77 BL Lacs and is the
largest complete XBL sample so far.

Recent models tried to explain the different evolutionary behaviour of
HBLs and LBLs by assuming that BL~Lacs\ start as LBLs and evolve into
HBLs as they grow older (Georganopoulos \& Marscher
\cite{georganopoulos}, Cavaliere \& D'Elia \cite{bms}). As described
by e.g. Padovani \& Urry the spectral energy distributions (SED)
of BL~Lacs\ are characterized by two components, both consisting of
beamed continuum emission from the plasma of the jets.  The first
component is synchrotron emission, peaking in the mm to far IR for
LBLs. The second component is inverse Compton (IC) emission peaking at
MeV energies. HBLs have SEDs peaking in the keV and in the GeV-TeV
band respectively. A decrease of power of the jets during the BL Lac
evolution would then be accompanied by an increase of the peak
frequencies and accordingly a transformation of the LBLs into HBLs
(Georganopoulos \& Marscher \cite{georganopoulos}).  This model is in
fact valid for the whole blazar class: BL Lacs in general show lower
power and beaming factors than the Flat Spectrum Radio Quasars
(FSRQs), as revealed by e.g. Madau et al. (\cite{madau}), Padovani
(\cite{padovani}), Ghisellini et al. (\cite{ghisellini3}). It
naturally explains the different evolution, which is slightly negative
for HBLs, slightly positive for LBLs, and clearly positive for the
FSRQ/blazar class.  This model explains the different types of BL Lac
objects only by different global intrinsic power (Maraschi \& Rovetti
\cite{maraschi}), and not by a different viewing angle. Nevertheless
different orientation is probably important as secondary effect
necessary to explain the large scatter of observed quantities.

The HRX-BL Lac sample contributes to the discussion with a large and
complete sample of X-ray selected BL Lac objects. Previous
studies (e.g. Fossati et al. \cite{fossati}) used a compilation of
different BL Lac surveys, like the X-ray selected EMSS (Stocke et
al. \cite{EMSS}, Rector et al. \cite{emssbllac}), the radio selected
1\,Jy BL Lac sample (Stickel et al. \cite{stickel}, Rector \& Stocke
\cite{rector}), and a FSRQ sample derived from the 2\,Jy radio sample
of Wall \& Peacock (\cite{2Jy}) to investigate the overall picture of
the blazar class, ranging from the FSRQs to the BL Lac objects. In
contrast to this the HRX-BL Lac survey is concentrating on a blazar
subclass, the HBL and IBLs, and is homogeneous in having the same
selection criteria for all objects, making it comparable with 
the REX-survey (Maccacaro et al. \cite{rex1}, Caccianiga et
al. \cite{rex2}), the DRXBS (Perlman et al. \cite{DXRBS}, Landt et
al. \cite{landt}), and the sedentary multifrequency BL Lac sample
(Giommi, Menna, \& Padovani \cite{sedentary}).

We will describe our selection method of BL Lac candidates in Section
\ref{sel} and the results of the identification process using
literature data and own observations in Section \ref{klass}. The
spectral energy distribution of the HRX-BL Lac\ sample is analyzed in
Section \ref{sed}, where we demonstrate that for the HBL class the
knowledge about the X-ray and optical flux is sufficient to determine
the peak frequency of the synchrotron branch. The spatial distribution
of the sample is described in Section \ref{raum}.  We conclude with a
discussion of the compatibility of the results from the HRX BL Lac
sample with recent studies.

Throughout the article a cosmology with $H_0 = 50 \rm \, km \,
sec^{-1} \, Mpc^{-1}$ and a deceleration parameter $q_0 = 0.5$,
assuming a Friedmann universe with $\Lambda = 0$, has been used.

\begin{table}
\caption[]{\label{hrxbl} Boundaries of the selection area.}
\begin{flushleft}
\begin{tabular}{ccr}
\multicolumn{2}{c}{Boundaries (J2000.0)} & Area \\
$\alpha$ & $\delta$ & [$\,{\rm deg}^2$] \\
\noalign{\smallskip}\hline\noalign{\smallskip}
 $7^{\rm h} \le \alpha < 8^{\rm h}$ & $30 \grad < \delta < 85 \grad$ & 426  \\
 $8^{\rm h} \le \alpha < 12^{\rm h}$ & $20 \grad < \delta < 85 \grad$ & 2248 \\
 $12^{\rm h} \le \alpha < 14^{\rm h}$ &  $20 \grad < \delta < 65 \grad$ & 970\\
 $14^{\rm h} \le \alpha \le 16^{\rm h}$ & $20 \grad < \delta < 85 \grad$ & 1124\\ 
\noalign{\smallskip}\hline
\end{tabular}
\end{flushleft}
\end{table}

\section{\label{sel} BL Lac candidate selection}
In the beginning the HRX-BL Lac sample originated from the Hamburg-RASS 
Bright X-ray AGN sample (HRX), which was created by identification of
the ROSAT All-Sky Survey (RASS) with the aid of objective prism plates of the
Hamburg Quasar Survey (HQS; Hagen et al. \cite{HQS}). The BL Lac subsample
was selected on an area of $1687 \,{\rm deg}^2$ with a count-rate limit
of $hcps \ge 0.075 \persec$  and on additional $1150 \,{\rm deg}^2$
with a limit of $hcps \ge 0.15 \persec$. This sample was analyzed by Bade et al. (\cite{bade})
and is referred here as the HRX-BL Lac {\it core sample}. It consists
of 39 BL Lacs, 34 of which are also part of
the present sample.

The fraction of BL Lacs in the HRX was $\sim 10 \%$. Therefore, an
increase of the sample size based on optical identification alone is
rather inefficient. This can be alleviated using radio information, as
all BL Lacs from the core sample were detected as radio sources in the
NRAO VLA Sky Survey (NVSS, Condon et al. \cite{NVSS}).  Also, to the
authors knowledge, all known BL Lac objects do have radio counterparts
down to the $\sim 2.5$ mJy level, which is similar to the detection
limit of the NVSS.  We concluded therefore that for the high X-ray
count-rates used we can include radio detection in the NVSS as
selection criterium without loosing BL Lac objects. As X-ray input we
used the ROSAT Bright Source Catalog (RASS-BSC; Voges et
al. \cite{RASSBSC}) with a count-rate limit ($0.5 - 2.0 \keV$) of
$hcps \ge 0.09 \persec$. We cross-correlated this catalogue with the
NVSS adopting an error circle of 30\arcsec\ around the X-ray position.
We extended the sky area studied to $4768 \,{\rm deg}^2$ 
encompassing the area of $2837 \,{\rm deg}^2$ studied by Bade et
al. (\cite{bade}), and we applied a unique limit of $hcps \ge 0.09 \persec$. 
The boundaries of the area are given in Table \ref{hrxbl}.

The cross-correlation yielded 223 matches between X-ray and radio
sources.  The complete list of these objects is given in Table
\ref{allcorr}
(page \pageref{allcorr}).
The coordinates listed are the X-ray
positions (J2000.0). More than $99.9 \%$ of the sources have a
positioning error $\Delta \le$ 25\arcsec\ (Voges et
al. \cite{RASSBSC}). The column ``Name'' lists alternative names to
the ROSAT designation, when available. Redshifts and classification
are taken from the NED or SIMBAD database or were determined on the
base of own follow-up observations. All objects,
for which we obtained own data are marked.

The cross correlation might be incomplete for lobe-dominated radio
sources, as in those cases the radio emission will consist of more
than one component offset from the X-ray position.
However, for none of the X-Ray BSC sources we found multiple radio sources within the
search radius, and as BL Lacs are core-dominated radio sources 
no selection biases are expected.

\section{\label{klass} Classification of BL Lac candidates}
\subsection{Results from NED and SIMBAD database queries}
The 223 candidates were classified in a two-step process. First we
searched for known optical counterparts in the 
NASA/IPAC Extragalactic Database (NED)\footnote{The NED is operated by 
the Jet Propulsion
Laboratory, California Institute of Technology, under contract with the 
National Aeronautics and Space Administration.} and in the SIMBAD\footnote{The SIMBAD Astronomical Database is operated by the Centre de Donn\'ees 
astronomiques de Strasbourg.} database. Special care was taken to avoid
confusion of BL Bac objects with normal galaxies and other types of AGNs.
A database entry of an object as ``galaxy'' without spectroscopic information 
was not accepted as an identification, because nearby BL Lac objects in 
elliptical galaxies might not have been recognized. For galaxies with 
redshift information and for objects with ``AGN'' or ``QSO'' identification
but without additional information (e.g. redshift)
the original literature was consulted before the object was dismissed
as BL Lac.

In total 101 objects could be classified this way.\footnote{at 
the time of writing this paper, already 205 of
the objects have a classification in the NED}
Of the remaining 
objects a few candidates were classifed as stars on the objective prism plates
of the HQS, and another few as obvious clusters of galaxies based on direct
plates and on the fact that these sources show extended X-ray emission. For all other candidates follow-up observations were
obtained. 

\begin{table}
\caption[]{Spectroscopic follow-up observation runs. 
The last column gives the number of objects observed.
Some of the objects were observed several times.}
\begin{tabular}{llcc} 
Telescope \& Instrument & Date & \#nights & $N^b$\\
\noalign{\smallskip}\hline\noalign{\smallskip}
3.5m Calar Alto (MOSCA) & March 1997    & 4 & 30\\
WHT / La Palma (ISIS)   & April 1997    & 2 & 19\\
3.5m Calar Alto (MOSCA) & Feb. 1998     & 6 & 89\\
3.5m Calar Alto (MOSCA) & Feb. 1999     & $\sim 1^{a}$ & 9\\
\noalign{\smallskip}\hline\noalign{\smallskip}
\end{tabular}

$^{a}$ morning and evening hours of three nights.\\
$^{b}$ number of objects observed in this observation run.
\label{hblobsruns}
\end{table}

\subsection{Observations of the remaining unclassified BL Lac candidates}
Spectroscopic observations to classify the remaining BL Lac candidates
and to determine their redshifts were made with the 3.5m telescope on
Calar Alto equipped with the multiobject spectrograph MOSCA and with
the 4.2m WHT on La Palma equipped with ISIS (Table \ref{hblobsruns}).
Most of the results from the 1997 observation runs have already been
presented in Bade et al. (\cite{bade}). 
For the classification with MOSCA, we used the G500 grism, which covers a
wavelength range $4250 - 8400$ \AA\, with a pixel-to-pixel resolution
of 12 \AA.  If necessary, additional spectra were taken with the G1000
and R1000 grisms to determine redshifts.  These  spectra
have a resolution of 6 \AA\, and cover the ranges
4400--6600 \AA\, and 5900--8000 \AA\,
respectively.  The spectra were reduced in a
standard way: bias subtraction, flat-field correction using morning
and evening skyflats, and response determination of the detector using
spectrophotometric standard stars.  BL Lac objects by definition have
no or very weak emission lines. Integration times of $\approx 1000
\dots 2000\, \rm sec$ were needed to detect the weak absorption lines
of the host galaxy, which is often out-shined by the non-thermal
continuum of the point-like central synchrotron source.  Most
observations were made under non-photometric conditions.

Optical photometry in the Johnson B band has been obtained for many of the optically faint BL Lacs with the Calar Alto 1.23m
telescope (Beckmann \cite{bldata}). Especially for several of the very
faint objects ($B > 20 \mag$) no reliable photometry was available
before.
For these objects we have now optical magnitudes with an error of
$\Delta B \le 0.1 \mag$. For the other objects the acquisition frames
of the 
spectroscopic runs have been used to determine a B magnitude, or
values from the literature have been taken. For the brighter objects
($B < 18 \mag$) also the HQS calibrated objective prism  plates
have been used, which have an error of $\Delta B \le 0.3 \mag$.

\subsection{Results from follow-up observations}
Follow-up observations were made of 117 objects, including the
unidentified candidates from the BSC/NVSS correlation and additionally
a number of objects, which were considered to be promising BL Lac
candidates, but did not match the selection criteria
described before (e.g. too low X-ray count rate).
In total we discovered 53 BL Lac objects 
according to the classification criteria, which are discussed in Section \ref{class}.
As a considerable number of the objects was discovered
independently by other groups in the meantime,
we are left over with 26 new BL Lacs: 11 from the BSC/NVSS correlation
and 15 from the additionally observed objects.

\subsubsection{The complete HRX BL Lac sample}
The 11 new BL Lac objects discovered among the objects from the
BSC/NVSS correlation are marked in Table \ref{allcorr} and their
spectra are included in Figure \ref{fig:spectra1}.  Based on our
spectra we could confirm or revise redshifts for several other BL Lac
objects. Five objects from this
correlation with other identifications than BL Lac and without NED or
SIMBAD entry so far, are marked in Table \ref{allcorr} in addition.

Summarizing, the optical identification of the 223 BSC/NVSS objects leads to
the following distribution of object classes within the radio/X-ray
correlation (Table \ref{hrxidentification}): 35\% are BL Lac objects,
36\% are other AGNs (QSO, Seyfert 1/2, FSRQs), 
11\% galaxies (including starburst galaxies and LINERs), 13\% cluster
of galaxies, and 4\% stars (including 2 supernova remnants). Only a
fraction of 1\% of the 223 candidates is yet not identified. 

\begin{table}
\caption[]{Summary of identifications of objects from the BSC/NVSS
correlation (cf. Table \ref{allcorr})}
\begin{tabular}{lrr}
object type & total number & fraction\\ 
\noalign{\smallskip}\hline\noalign{\smallskip}
BL Lac      & 77           & 34.5\%\\
Seyfert 1   & 65           & 29.1\%\\
Seyfert 2   &  8           &  3.6\%\\
Quasar      &  6           &  2.7\%\\
blazar      &  2           &  0.9\%\\
LINER       &  5           &  2.2\%\\
Galaxy cluster & 29        & 13.0\%\\
Galaxies    & 19           &  8.5\%\\
Stars       &  8           &  3.6\%\\
SNR         &  2           &  0.9\%\\
Unidentified & 2           &  0.9\%\\
\noalign{\smallskip}\hline\noalign{\smallskip}
Total        & 223         & \\
\noalign{\smallskip}\hline\noalign{\smallskip}
\end{tabular}
\label{hrxidentification}
\end{table}

The 77 BL Lacs from the BSC/NVSS correlation are called the {\it
complete HRX BL Lac sample}.  In comparison to the EMSS BL Lac sample,
this sample probes a population of objects with lower $\aro$ and
$\aox$ values and contains therefore more radio quiet and stronger
X-ray dominated objects. The HRX-BL Lac sample is the largest complete
sample of X-ray selected BL Lac objects. Table \ref{samples} compares
the HRX BL Lac sample with four other X-ray selected BL Lac
Surveys: the EMSS based sample (Rector et al. \cite{emssbllac}), the
sample by Laurent-Muehleisen et al. (\cite{RGB}) based on the
correlation of the RASS with the Green Bank radio survey, the REX
survey using the NVSS in combination with the sources found in the
ROSAT pointed observations (Caccianiga et al. \cite{rex2}), and the
DXRBS (Perlman et al. \cite{DXRBS}), which uses the ROSAT data base
WGACAT and PMN/NVSS radio data.

\begin{table*}
\caption[]{Properties of the Hamburg BL Lac samples in comparison to other recent samples}
\begin{tabular}{lcclll}
 sample & Reference & number of & X-ray & radio & optical\\
        &           & objects   & limit & limit & limit  \\ 
\noalign{\smallskip}\hline\noalign{\smallskip}
HRX core sample & Bade et al. 1998 & 39 & $0.075/0.15 \persec$ $^{a,g)}$ & - & - \\
HRX-BL Lac & this work & 77 & $0.09 \persec$ $^{a,g)}$ & $2.5 \mJy^{b)}$ & -\\
RGB & Laurent-Muehleisen  & 127 & $0.05 \persec$ $^{a,c)}$ & $15 \dots 24 \mJy^{d)}$ & $18.5 \mag^{e)}$\\
RGB complete & et al. 1999 & 33 &  $0.05 \persec$ $^{c)} $ & $15 \dots 24 \mJy^{d)}$ & $18.0 \mag^{e)}$\\
EMSS & Rector et al. \cite{emssbllac} & 41 & $2 \times 10^{-13}$ $^{f)}$ & - & -\\
REX  & Caccianiga et al. \cite{rex3} & 55 & $4 \times 10^{-13}$ $^{g)}$ & $2.5 \mJy^{b)}$ &  $B \le 20.5 \mag$\\   
DXRBS & Padovani \cite{padovanidxrbs} & 30  & few $\times 10^{-14}$ $^{c)}$ & $\sim 50 \mJy$ & -\\
\noalign{\smallskip}\hline\noalign{\smallskip}
\end{tabular}\\
$^{a)}$ ROSAT All Sky Survey count rate limit\\
$^{b)}$ NVSS radio flux limit at $1.4 \GHz$\\
$^{c)}$ Full ($0.1 - 2.4 \keV$) PSPC energy band.\\
$^{d)}$ GB catalog flux limit at $5 \GHz$\\
$^{e)}$ $O$ magnitude determined from POSS-I photographic plates\\
$^{f)}$ EINSTEIN IPC ($0.3 - 3.5 \keV$) flux limit in $[\ecs]$\\
$^{g)}$ hard ($0.5 - 2 \keV$) PSPC energy band.
\label{samples}
\end{table*}

\subsubsection{The extended HRX BL Lac sample}
Among the BL Lac candidates observed additionally we could confirm
another 27 BL Lac objects, from which 15 did not appear in the
literature so far. Together with the 77 BL Lacs from the complete
sample they form the {\it extended HRX BL Lac sample} comprising 104
objects. The spectra of the 15 new BL Lacs are presented in Figure
\ref{fig:spectra1} together with the 12 new BL Lacs from the complete
sample.  The spectra of all other objects including those from objects
not identified as BL Lacs will be accessible
on the web\footnote{http://www.hs.uni-hamburg.de/bllac.html}.

The properties of the 104 BL Lacs of the extended sample are presented
in Table \ref{hblsample1}
(page \pageref{hblsample1}). The BL Lacs discovered
additionally are marked by an asterisk and the new BL Lacs are labeled
by ``new''.  This Table lists the object names, the NVSS radio
coordinates (J2000.0), redshifts, ROSAT PSPC (0.5 - 2.0 keV) X-ray
fluxes in $10^{-12}\, \ecs$, 1.4 GHz radio fluxes in mJy from the NVSS
radio catalogue, B magnitudes, K magnitudes, and the calcium break
index. The radio positions have an error of
less than 5\arcsec\ (for
the faintest objects) and are therefore considerably more accurate
than the X-ray positions given in Table \ref{allcorr}.
 
The RASS-BSC fluxes have been computed by using the count rate and a
single-power law with free fitted absorption $N_H$. The spectral slope
and $N_H$ are determined by the hardness ratios, a method
described by Schartel (\cite{schartel}). The hardness
ratio is defined as $HR = (H-S)/(H+S)$ with $H$ and $S$ being the
number of counts in the hard and soft energy bands; typically two
ratios are computed: $HR1$ with energy ranges $S = 0.1 - 0.4 \rm \,
keV$ and $H = 0.5 - 2.0 \rm \, keV$, and $HR2$ with $S = 0.5 - 0.9 \rm
\, keV$ and $H = 1.0 - 2.0 \rm \, keV$ (Voges et al. \cite{RASSBSC}). The
values for the hardness ratios range by definition from $+1$ for
extremely hard to $-1$ for very soft X-ray spectra. The error estimate
for the $N_{\rm H}$ and $\alpha_{\rm X}$ values is based on the
hardness ratios
only, not on the photon spectrum itself. Therefore this method does
not give $\chi^2$ values, but is able to determine $68 \% \, (1
\sigma)$ errors. This is done by exploring the hardness-ratio,
spectral slope, and $N_{\rm H}$ parameter space, determining the $1 \sigma$
region within it for a given set of parameter components.

The near infrared data are taken from the Two-Micron All-Sky
Survey (2MASS, Skrutskie et al. \cite{2MASSa}, Stiening et al. \cite{2MASSb}).
In Table \ref{hblsample1} only the $K$-magnitude is listed, but for the
analysis we also used $J$ and $H$ from the 2MASS.

The calcium break index (Ca-break) is defined as follows (Dressler \&
Shectman \cite{cabreak}):
\begin{equation}
Ca-break[\%] = 100 \cdot \frac{f_{\rm upper} - f_{\rm lower}}{f_{\rm upper}}
\end{equation}
with $f_{\rm upper}$ and $f_{\rm lower}$ being the mean fluxes measured in the 3750
\AA\, $< \lambda <$ 3950 \AA\, and 4050 \AA\, $< \lambda <$ 4250 \AA\, objects
rest frame band respectively. 
The 
measurement of the break was possible only if the redshift was known and
if the break was within the spectral range covered by the
observation. Therefore break values are available for 30 of the HRX-BL Lac\ only. In
seven cases the break value is negative. This is in most cases
  due to a low signal to noise of the spectra. Only for one
object (1RXS J111706.3+201410) the negative calcium break index is not
consistent with a value of 0\% and we assume that here the underlying
power law of the jet emission outshines the host galaxy, so that the
measured negative 'break value' is in fact based on the synchrotron component.

\begin{figure}
\begin{center}
\epsfxsize=7.0cm
\epsfbox{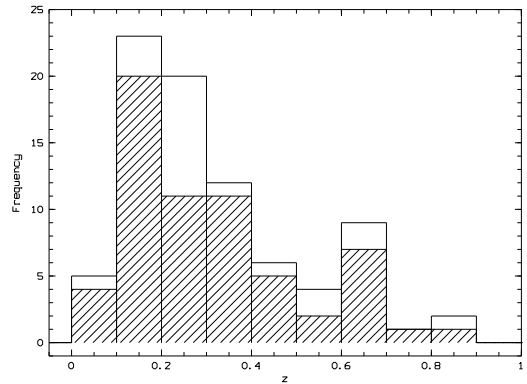}
\caption[]{\label{fig:hblz}Distribution of redshifts in the extended HRX-BL
  Lac sample. The hatched  part refers to the {\it complete sample}.} 
\end{center}
\end{figure}

Figure~\ref{fig:hblz} shows the redshift distribution of the HRX-BL
 Lac extended and complete samples.
The mean redshift for the {\it complete} and {\it extended sample} are
$\bar{z} = 0.31$ and $\bar{z} = 0.34$,
respectively. We note that in comparison to the {\it core sample} no
new BL Lacs with $z > 0.7$ were found, which contribute to the
{\em complete sample}.

\subsection{\label{class} BL Lac classification criteria}

The characterizing feature of BL Lac spectra in the optical is the presence of
a non-thermal continuum which is well described by
a single power law. A second component is the emission of the host
galaxy, which contributes absorption features in addition to continuum
emission.  If the BL Lac itself shows no emission lines at all, 
redshift determination is only possible by identifying these
absorption features. The host galaxies are in majority giant
elliptical galaxies (e.g. Urry et al. \cite{urryhost}), having strong absorption
features caused by the stellar content.

Expected absorption features in the optical, which can be used for
redshift determination, have already been discussed in detail by Bade
et al. (\cite{bade}). The most prominent feature in the spectra of
elliptical galaxies is the so-called ``calcium break'' at 4000
\AA. Its strength is given by the calcium break index, as defined
before.

Most of the AGN with emission lines found in the radio/X-ray
correlation are Seyfert type galaxies or LINER (see Table \ref{hrxidentification}). These AGN do not show a calcium break.
For the other objects the strength of the
calcium break can be used to distinguish between normal elliptical
galaxies and BL Lac objects. For the former, this contrast is $\ge
40\%$ with the higher flux to the red side of the break. Our criteria
to classify BL Lac objects were defined by Bade et al.  (\cite{bade})
for the core sample and are spectroscopically similar to those applied
to the Einstein Medium-Sensitivity Survey (EMSS; Stocke et
al. \cite{EMSS}).  However, we relaxed the upper limit for the
strength of the calcium break index from 25\% to now 40\% when other
properties of the object were consistent with a BL Lac
classification. This follows the findings of previous studies
(March\~a et al. \cite{marcha}; Laurent-Muehleisen et al. \cite{RGB};
Rector et al. \cite{emssbllac}) that there exist galaxies with
strengths $25\% <$ Ca-break $< 40\%$, which fulfill all other
selection criteria for BL Lac objects. Explicitly the selection
criteria are now:
\begin{itemize}
\item No emission lines with W$_{\lambda} > 5$ \AA\
\item The contrast of the Ca II break from the host galaxy must be less
than 40\%.
\end{itemize}

With respect to the first criterium no misclassifications are expected
as there were no objects found within the BSC/NVSS
correlation with weak emission lines and equivalent widths of several
10 \AA.

\begin{figure}
\begin{center}
\epsfxsize=7.0cm
\epsfbox{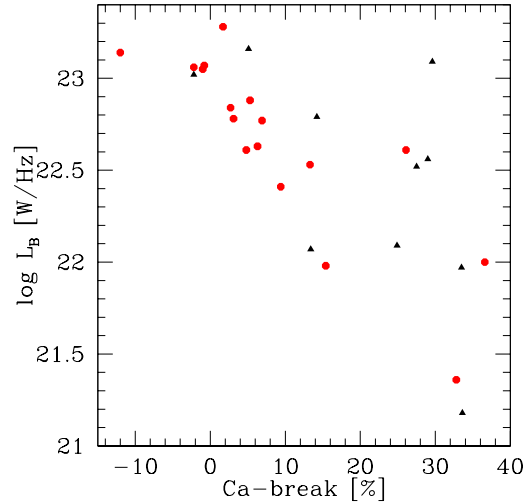}
\caption[]{\label{fig:hblbreaklum}Strength of the calcium break versus
  monochromatic luminosity L$_B$ in the optical B-band. Circles refer to the {\em complete sample} while triangles mark additional objects found within the course of the work.}
\end{center}
\end{figure}

Borderline cases are more likely with respect to the calcium break index,
because the transition between non-active elliptical galaxies and BL
Lacs is smooth. This is clearly shown in Figure \ref{fig:hblbreaklum},
in which our measured  break strength is plotted vs. the optical luminosity
L$_B$, as derived in Section \ref{lumino}. Both quantities are 
correlated and almost evenly distributed up to Ca-break $\sim$40\%. 

The observed correlation might be affected by a varying fraction of host
galaxy light included in the spectra. In nearby objects the BL Lac host
galaxy might not have been fully covered by the slit and therefore
the calcium break strength could have been underestimated. However, as the
low-redshift objects are mainly the less luminous ones, this effect
cannot explain the decreasing strength of the calcium break with
increasing luminosities.

This correlation is not only seen in the optical domain, but is also
present if we use radio, near infrared or X-ray luminosity
instead. In all wavelength regions from the optical to the
X-rays the correlation between emitted luminosity and break strength
is significant. 
Therefore we would like to
stress the point that the observed correlations are  not due to
observational selection effects.

Misclassifications might have been occurred also due to large
errors for the measured break strengths in some of our spectra with
low signal to noise ratio. However for all objects except three of
the HRX-BL Lac {\em complete sample} the break strengths are $<$25\%,
making a misclassification unlikely. The three objects with a calcium
break strength in the range $25\% < Ca-break < 40\%$ are 1ES 0927+500,
1RXS 114754.9+220548, and 1RXS 151040.8+333515 (cf. Table
\ref{hblsample1}) and they were included in the sample, because they
fulfill other BL Lac properties, for example strong polarization ($P >
6\%$) in the NVSS.

\section{\label{sed} Spectral Energy Distribution}
To study the spectral energy distribution (SED) of the HRX-BL Lac objects,
overall spectral indices were calculated to derive general
correlations within the sample. Throughout this Section
the {\em extended sample} is analyzed.
Ledden and O'Dell (\cite{indices}) defined
the overall spectral index between two frequencies:
\begin{equation}
\alpha_{1/2} = - \frac{\log (f_{1} / f_{2})}{\log (\nu_{1} / \nu_{2})}  
\end{equation}
Here $f_{1}$ and $f_{2}$ are the fluxes at two frequencies $\nu_{1}$ and
$\nu_{2}$. As reference frequencies we used 1.4 GHz
in the radio ($\lambda \simeq 21 \, \rm cm$), 4400 \AA\, in the
optical ($\sim B$), and $1 \keV$ ($\lambda \simeq 12.4$ \AA) in the
X-ray region to derive the optical-X-ray $\alpha_{OX}$, the
radio-X-ray $\alpha_{RX}$, and the radio-optical $\alpha_{RO}$ spectral index.

To compare these indices with those from the literature, shifts
due to the use of different reference energies have to be taken into
account.  It can be shown that these shifts are small as long as the
spectral shape within each band can be approximated by a single power
law and the spectrum is not curved.
Because the radio spectra are
flat ($\alpha_R = 0$), the flux does not change when different reference frequencies 
are chosen in the radio domain. But by increasing the radio reference 
frequency, the $\alpha_{\rm RX}$ and $\alpha_{\rm RO}$ indices steepen.
For example, if the reference
frequency is changed from 1.4 to 5 GHz the radio-X-ray index changes
by 6\%: $\alpha_{RX} (5 \GHz,1 \keV) \simeq 1.06 \times \alpha_{RX}
(1.4 \GHz,1 \keV)$. If our spectral indices are compared with those
using a larger X-ray reference energy, similar values for $\aox$ and
$\arx$ are expected. Because of $f_\nu \propto \nu^{-\alpha}$, the
expected flux at a higher energy is lower and the flux ratios
increase. At the same time however, the frequency interval increases by about
the same factor, if we assume $\alpha_E = 1$, which is a good
approximation for the mean X-ray spectral energy index of BL Lac
objects 
The same reasoning applies for the optical region, where
$\alpha_E \lae 1$, and larger changes of the relevant indices are not
expected.

The HRX-BL Lac sample shows typical values for the mean overall spectral
indices: $<\alpha_{OX}> = 0.94 \pm 0.23$,  $<\alpha_{RX}> = 0.55 \pm
0.08$, $<\alpha_{RO}> = 0.37 \pm 0.09$,  if compared to 
Wolter et al. (\cite{saxpap1}), Laurent-Muehleisen et
al. (\cite{RGB}), and Beckmann et al. (\cite{saxpap}). 

\begin{figure}
\begin{center}
\epsfxsize=8.0cm
\epsfbox{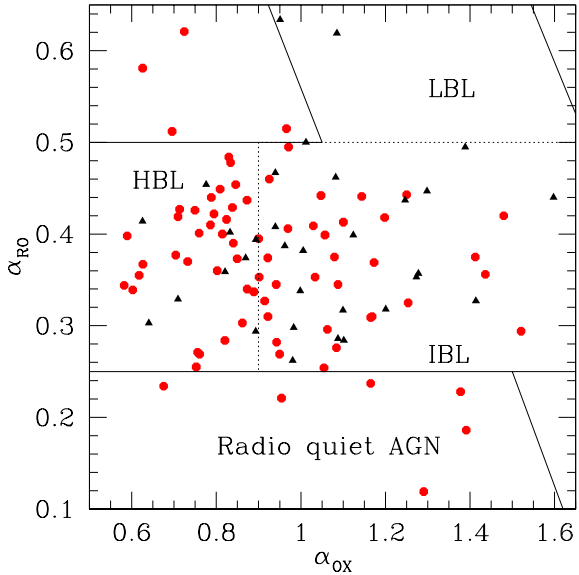}
\caption[]{\label{fig:hblaoxaro}The $\aox - \aro$ plane covered by the
  HRX-BL Lac objects. The points refer to the {\it complete sample}, the
  triangles mark additional objects found within the course of the
  work. Objects with $\aro < 0.2$ are called radio quiet.}
\end{center}
\end{figure}
The region
in the $\aox - \aro$ plane, which is covered by the HRX-BL Lac
sample, is shown in Figure \ref{fig:hblaoxaro}. 
The
center of the area covered by this sample is similar to that of the
EMSS BL~Lacs (see Padovani \& Giommi \cite{emss}) though a larger range in
$\aox$ and $\aro$ is covered.

\subsection{Peak frequency}

In order to get a more physical description of the spectral energy
distribution of the BL Lac objects, we used a simple model to fit the
synchrotron branch of the BL Lac. This has the advantage of describing
the SED with one parameter (the peak frequency) instead of a set of
three parameters ($\aox$, $\aro$, and $\arx$). It has been shown
by several authors that the synchrotron branch of the BL Lac SED is
well approximated by a parabolic fit in the $\log \nu - \log \nu
f_\nu$ plane
(cf. Landau et al. \cite{landau}, Comastri et
al. \cite{pspcreduc}, Sambruna et al. \cite{sambruna}, Fossati et
al. \cite{fossati}). In this way the peak position  ($\nu_{\rm peak}$), the
total luminosity and the total flux of the synchrotron 
emission can be derived. We chose the parameterization using
fluxes $\log \nu f_\nu = a \cdot (\log \nu)^2 + b \cdot \log \nu +
c$. Using luminosities instead of fluxes would change the absolute
constant {\em c} only, leaving the position of the peak frequency unaffected.
 
\begin{figure}
\begin{center}
\epsfxsize=8.0cm
\epsfbox{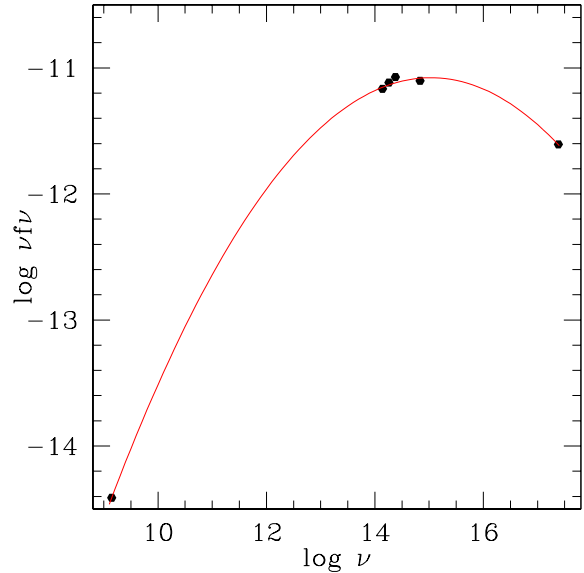}
\caption[]{\label{fig:examplesed}Parabolic fit to the data of B2 0912+29.}
\end{center}
\end{figure}
  
\begin{figure}
\begin{center}
\epsfxsize=8.0cm
\epsfbox{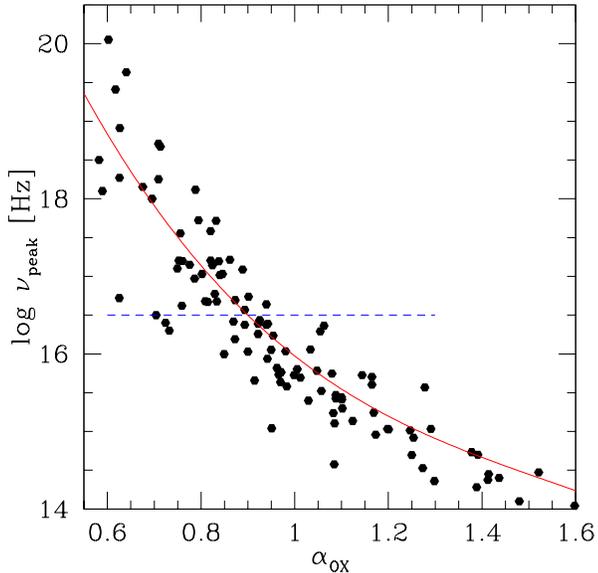}
\caption[]{\label{fig:hblaoxpeak}Logarithm of the peak frequency vs. $\aox$. The relation was approximated by a polynomial of third degree. The horizontal line marks the distinction between HBLs (above the line) and IBLs used in this paper.}
\end{center}
\end{figure}

If only three data points were given (one in the radio, optical, and
X-ray band), the parabola was definite. When more than three data
points were available (i.e. the $K$, $H$, and $J$ near infrared
measurements from the 2MASS), a $\chi^2$ minimization was used to
determine best fit parameters. In principle, also the spectral slope
in the X-ray band could be used to constrain the fit further. Because 
of the large uncertainties involved deriving these slopes from 
hardness ratio in the RASS, they were used mostly for consistency checks.
Only in cases, where the  parabolic fit
resulted in peak frequencies above the highest energies observed,
i.e. for objects with $\log \nu_{\rm peak}$ above 2 keV and 
steep X-ray spectra, the slopes were taken to account. 
An example for a parabolic fit is shown in Figure
\ref{fig:examplesed}. $\nu_{\rm peak}$ is sensitive for the
$f_x/f_{opt}$-relation, and is therefore strongly correlated with
$\aox$. This is shown in Figure \ref{fig:hblaoxpeak}.
The relation can be approximated by a polynomial of third
degree. Using an F-test, parabolic fits of higher degree
 gave no improvements.
Thus the peak frequency can also be determined, in the case  no radio
data is available by applying
\begin{equation}
\log \nu_{\rm peak} = -3.0 \cdot \aox^3 + 13.8 \cdot \aox^2 - 23.3 \cdot \aox + 28.5 (\pm 0.51)
\label{peakaoxequation}
\end{equation}
The standard error ($\sigma = 0.51$) is based on the deviation of the data
points from the fit in Figure \ref{fig:hblaoxpeak}.  This result is comparable to that found by Fossati et al. (\cite{fossati}) when studying the dependency of $\alpha_{RO}$ and $\alpha_{RX}$ on the peak frequency.

\subsection{\label{lumino} Correlation of luminosities with the SED}

A set of physical parameters which are correlated to the peak
frequency are the luminosities in the different wavelength regions.  
To compute luminosities for all
objects, the unknown redshifts were set to $z = 0.3$ which is the mean
value for the HRX-BL Lac sample. 
While the luminosities  $L_R$ in the radio, $L_K$ in the near
infrared, and $L_B$ in the optical region are decreasing with
increasing peak frequency, the situation at X-ray energies is the
other way round
(as reported also by e.g. Mei, Zhang, \& Jiang \cite{Mei}, Beckmann \cite{beckmann}).

The details about the correlation analysis are listed in Table
\ref{hblpeaklumtab}, including the confidence level of the correlations.
\begin{table*}
\caption[]{Correlation of luminosity with peak frequency in the extended HRX-BL Lac sample}
\label{hblpeaklumtab}
\begin{tabular}{llll}
region & $r_{xy}$ Pearson     & confidence level & linear regression$^a$ \\
       & coefficient & of correlation &\\
\hline
radio $(1.4 \GHz)$   & -0.23 & $> 97 \%$   & $\log L_R = -0.09 \cdot \log \nu_{\rm peak} + 26.4$ \\ 
near IR (K-band)     & -0.28 & $> 95 \%^b$ & $\log L_K = -0.14 \cdot \log \nu_{\rm peak} + 25.9$ \\ 
optical (B-band)     & -0.37 & $> 99.9 \%$ & $\log L_B = -0.13 \cdot \log \nu_{\rm peak} + 25.1$ \\ 
X-ray   $(1 \keV)$   & +0.51 & $> 99.9 \%$ & $\log L_{\rm X} = +0.19 \cdot \log \nu_{\rm peak} + 17.3$ \\[0.2cm] 
total (radio -- X-ray) & -0.12 &  & $\log L_{sync} = -0.04 \cdot \log \nu_{\rm peak} + 22.0$ \\
\noalign{\smallskip}\hline
\end{tabular}

$^a$ Luminosities in $[\whz]$\\
$^b$ The lower confidence level results from the lower number of objects (52) with known K-band magnitudes. The other correlations are using the 104 BL Lacs of the extended sample. 
\end{table*}

The total luminosity L$_{sync}$ within the synchrotron branch has been
derived by integrating the spectral energy distribution between the
radio and the X-ray band. This is a reasonable approximation as
long as the peak frequency is below 1 keV ($\log \nu = 17.4$), but
systematically underestimates L$_{sync}$
if the peak frequency is
shifted beyond $1 \keV$.  The relation of peak frequency with the
total luminosities does not show a clear correlation. 

\section{\label{raum} Distribution in space}
\subsection{$V_{\rm e}/V_{\rm a}$-test}
Redshifts are available for 64 (83\%) 
of the 77 BL Lac objects which
form the {\it complete sample}. Therefore it is possible to determine a
luminosity function for the HRX-BL Lacs and to study the evolution by
application of an $V_{\rm e}/V_{\rm a}$-test. 
The direct images of the BL Lacs
without redshift determination show point-like structure, and most of
them have optical spectra consistent with high redshifts
($z>0.5$). The difficulty in determining redshifts for them indicate
that these objects are highly core dominated with the host galaxy
outshined by the BL Lac core. This implies that the optical luminosity of these objects should be quite high.

The $V_{\rm e}/V_{\rm a}$-test is a simple method developed by Avni \& Bahcall
(\cite{veva}) based on the $V/V_{\rm max}$ test of Schmidt
(\cite{schmidt}).  $V_{\rm e}$ stands for the volume, which is enclosed by
the object, and $V_{\rm a}$ is the accessible volume, in which the object
could have been found (e.g. due to a flux limit of a survey).
Avni \& Bahcall showed that different survey areas with different flux limits in various energy
bands can be combined by the $V_{\rm e}/V_{\rm a}$-test.  In the case of no
evolution $ \langle V_{\rm e}/V_{\rm a} \rangle = 0.5$ is expected and following
Avni \& Bahcall (\cite{veva}) the error $\sigma_m(n)$ for a given
mean value $\langle m \rangle = \langle V_{\rm e}/V_{\rm a} \rangle$ based on $n$ objects is:
\begin{equation}
\sigma_m(n) = \sqrt{\frac{1/3 - \langle m \rangle + \langle m \rangle^2}{n}}
\end{equation}
We computed the accessible volume $V_{a,i}$ for each object by
applying the survey limits. In most cases this volume is determined by
the X-ray flux limit, only $\sim 10\%$ of the objects have a smaller
$V_{a,i}$ for the radio data, due to the radio flux limit of $2.5
\mJy$.

Applied to the {\em complete sample} the test yields $\langle V_{\rm
e}/V_{\rm a} \rangle = 0.42 \pm 0.04$. This result shows that HBLs
have been less numerous and/or less luminous in the past, but the
significance is only $2 \sigma$. The negative evolution of X-ray
selected BL Lac objects has been reported several times before. We
also performed a K-S test in order to determine the probability of
uniform $V_{\rm e}/V_{\rm a}$ distribution, which would mean no
evolution. For the whole HRX-BL Lac sample the probability of no
evolution is rather small ($3.5 \%$).

Thanks to the large number of objects with known redshifts within the
\mbox{HRX-BL Lac} sample it is possible to examine dependencies of the
evolution on other parameters, like the overall spectral indices.  
A division into two groups (more and less X-ray
dominated objects) according to $\aox$ was already made by Bade et
al. (\cite{bade}) for the {\em core sample} and resulted in a lower
$\langle V_{\rm e}/V_{\rm a} \rangle$= $0.34\pm0.06$ for the HBLs ($\aox < 0.9$)
than for the IBLs within the sample. The $\langle V_{\rm e}/V_{\rm a} \rangle$ =
$0.48\pm0.08$ for IBLs was even consistent with no evolution.
Dividing the HRX-BL Lac sample accordingly we now get 
for the HBLs ($\aox < 0.9$) $\langle V_{\rm e}/V_{\rm a} \rangle = 0.45 \pm 0.05$
(N=34) and for the IBLs $\langle V_{\rm e}/V_{\rm a} \rangle = 0.40 \pm 0.06$
(N=30). The difference between the two groups has practically vanished,
and we are thus
not able to confirm the different types of
evolution for the HBLs and the IBLs. But still there are 13
objects within the HRX-BL Lac sample without known redshift, and nearly all
of them are IBLs. Including them into the $V_{\rm e}/V_{\rm a}$-test by
assigning them either the mean redshift of our sample ($z=0.3$)
or a high redshift ($z=0.7$) does
not change the mean $V_{\rm e}/V_{\rm a}$
values significantly.
The results of the different  $V_{\rm e}/V_{\rm a}$-tests are shown 
in Table~\ref{vvmax}. 
Assigning even higher redshifts
would increase the $V_{\rm e}/V_{\rm a}$ for the IBLs, but we consider this 
unlikely, as the luminosities would then become exceptionally high.
For example in  0716+714, PG 1246+586, or PG 1437+398 the X-ray luminosities would exceed values of $L_{\rm X} = 10^{46} \; \rm erg \; s^{-1}$ in the $0.5 - 2.0 \; \rm keV$ range.

\begin{table}
\caption[]{\label{vvmax} Results from the $V_{\rm e}/V_{\rm a}$-tests for 
the HRX-BL Lac {\em complete sample} }
\begin{flushleft}
\begin{tabular}{lcllc}
selection & unknown $z$ & $N^a$ & $\langle V_{\rm e}/V_{\rm a} \rangle$& K-S$^b \, [\%]$ \\
 & set to & & \\
\noalign{\smallskip}\hline\noalign{\smallskip}
all (known $z$) & -   & 64 & $0.42 \pm 0.04$  &3.5\\
all                & 0.3 & 77 & $0.44 \pm 0.03$  &5.3\\
all                & 0.7 & 77 & $0.46 \pm 0.03$  &5.3\\
HBLs (known $z$) & -   & 34 & $0.45 \pm 0.05$ &24.0\\
all HBLs            & 0.3 & 36 & $0.48 \pm 0.05$ &46.1\\
all HBLs            & 0.7 & 36 & $0.48 \pm 0.05$ &46.1\\
IBLs (known $z$) & -   & 30 & $0.40 \pm 0.06$ &14.0\\
all IBLs            & 0.3 & 41 & $0.41 \pm 0.05$ &10.7\\
all IBLs            & 0.7 & 41 & $0.43 \pm 0.05$ &10.7\\
\noalign{\smallskip}\hline\noalign{\smallskip}
\end{tabular}
$^a$ number of objects used for this test\\
$^b$ K-S test probability that the $V_{\rm e}/V_{\rm a}$ values have a uniform distribution in the $[0...1]$ interval (probability for no evolution). 
\end{flushleft}
\end{table}

\begin{table}
\caption[]{\label{vvmax2} Results from the $V_{\rm e}/V_{\rm a}$-tests for 
comparable investigations}
\begin{flushleft}
\begin{tabular}{lclll}
survey & selection & unknown $z$ & $N^a$ & $\langle V_{\rm e}/V_{\rm a} \rangle$ \\
\noalign{\smallskip}\hline\noalign{\smallskip}
REX & total & 0.27 & 55 & $0.48 \pm 0.04$\\
REX & HBL   & 0.27 & 22 & $0.49 \pm 0.06$\\
sedentary & total & 0.25 & 155 & $0.42 \pm 0.02$\\
DXRBS     & all BL Lacs & 0.40 & 30 & $0.57 \pm 0.05$\\ 
DXRBS     & HBL & 0.40 & 11 & $0.65 \pm 0.09$\\ 
DXRBS     & LBL & 0.40 & 19 & $0.52 \pm 0.07$\\ 
\noalign{\smallskip}\hline\noalign{\smallskip}
\end{tabular}
$^a$ number of objects used for this test
\end{flushleft}
\end{table}

We conclude therefore that the HRX sample shows no difference in
evolution for HBLs and IBLs. The results presented here are in good
agreement with recent other investigations on the evolutionary
behaviour of BL Lac objects, as shown in Table~\ref{vvmax}. Except the
sedentary survey (Giommi, Menna, \& Padovani \cite{sedentary}) none of
the studies could confirm the highly significant negative evolution
found e.g. by Bade et al. (\cite{bade} for the HRX-BL Lac {\em core
sample} or by Wolter et al. (\cite{wolter94}) for the EMSS BL
Lacs. The best sample to be compared with should be the REX survey,
which also uses the combination of RASS and NVSS data, although going
to lower X-ray flux limits while using only the are of the PSPC
pointed observation. The REX has also a mean redshift of $z = 0.3$ and
the $\langle V_{\rm e}/V_{\rm a} \rangle$ are within one sigma when
compared to the HRX-BL Lac sample.

\subsection{X-ray Luminosity functions}
For the $V_{\rm e}/V_{\rm a}$-test the knowledge of the redshifts
is of minor importance. However, the lumminosity function which
defines the space density at a given object luminosity can only be
derived when having a complete sample with known distances. Based on
this function we wish to estimate the fraction of AGN which appear to be
BL Lac objects.  To determine the cumulative luminosity function
(CLF), one has to count all objects within a {\it complete sample}
above a given luminosity, and divide this number by the volume $V_{\rm
a}$ which has been surveyed for these objects. 
We follow here the procedure described by Marshall (\cite{marshall}) to
derive the space density and the corresponding errors.
The survey area of the HRX-BL Lac {\it complete sample} is $4768
\,{\rm deg}^2$ (Table \ref{hrxbl}) . Because the fraction of objects
without known redshift is 17\% the effective area which is used to
compute the luminosity function is decreased by this fraction to $3959
\,{\rm deg}^2$. This implies that the redshift distribution of the
missing objects is the same as for the rest. As discussed before, this
assumption might be incorrect, as we expect many of them having rather
high redshifts. The effect of different evolution for high and low
redshift objects, described in the
next paragraph, would be even stronger in this case.

\begin{figure}
\begin{center} \epsfxsize=8.0cm \epsfbox{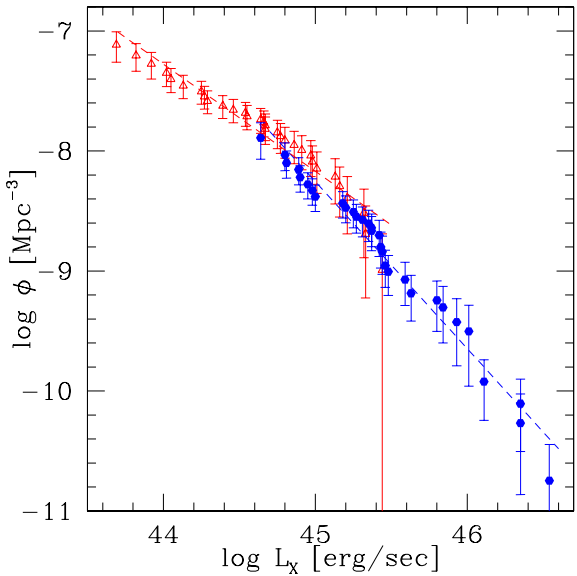}
\caption[]{\label{fig:hblcumlhz} 
Cumulative luminosity function of the two subsamples with $z > 0.272$ (circles) and $z < 0.272$ (open triangles).}
\end{center}
\end{figure}

The {\em complete sample}
is large enough to divide it into a high
redshift and a low redshift bin in order to examine possible
differences in their CLF. The dividing value was set to the median of
the HRX-BL Lac sample $z_{median} = 0.272$. To derive high and low redshift
CLFs the accessible volume $V_{a,i}$ for the objects with $z < 0.272$
has been restricted to $z = 0.272$ whenever $z_{max,i} > 0.272$. For
the high redshift objects the accessible volume was computed from $z =
0.272$ up to $z_{max,i}$. The resulting two cumulative luminosity
functions are shown in 
Figure \ref{fig:hblcumlhz}.

There seem to be differences between the high and low redshift
CLF. The slope of the low redshift CLF is flatter. A linear
regression gives a slope of $-0.9$ while for $z > 0.272$ the slope is
$-1.4$. But in the overlapping regime at $L_{\rm X}(0.5 - 2.0 \keV) \sim
10^{45}$ the luminosity functions show similar slopes. 

\begin{figure*}[t]
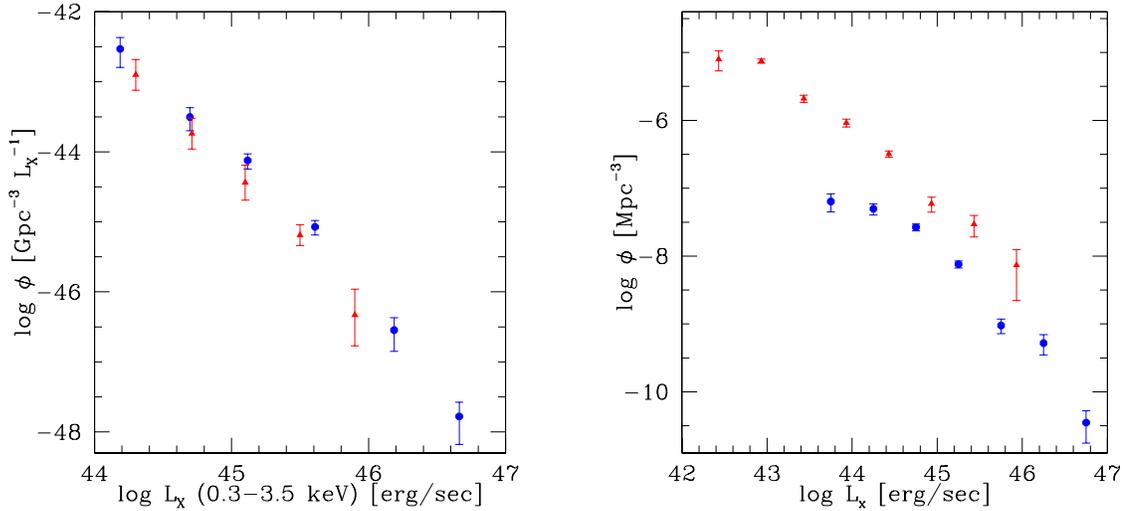

 \begin{minipage}[t]{15cm}
 \begin{minipage}[t]{7.0cm}
  \begin{flushleft}
  \psfig{figure=2548.f7,width=7.0cm}
  \end{flushleft}
 \end{minipage}
 \hfill
 \begin{minipage}[t]{7.0cm}
  \begin{flushright}
  \psfig{figure=2548.f8,width=7.0cm}
  \end{flushright}
 \end{minipage}
\end{minipage}
\caption[]{\label{fig:hbldiffcompare} Left panel: The differential X-ray luminosity function of the HRX-BL~Lac {\it complete sample} (circles) in comparison to EMSS BL~Lacs (triangles; Padovani \& Giommi \cite{emss}). The X-ray data of the HRX-BL~Lac objects have been extrapolated to the EINSTEIN IPC energy band assuming a spectral slope of $\alpha_{\rm X} = 1$. Right panel: Comparison of the X-ray luminosity function of RASS selected AGNs from the ROSAC sample
 (triangles; Tesch \cite{tesch}) with HRX-BL Lacs (circles). The
 density of BL Lacs is $\sim 10$ times lower than for all
 AGNs.}
\end{figure*}

The left panel of Figure
\ref{fig:hbldiffcompare} shows the comparison of the HRX-BL~Lac {\it
complete sample} X-ray luminosity function with the results from the
EMSS BL~Lac sample (Wolter et al. \cite{wolter94}, Padovani \& Giommi
\cite{emss}). The expected luminosities of the HRX-BL~Lacs within the
EINSTEIN IPC energy band ($0.3 - 3.5 \keV$) were calculated assuming a
spectral slope of $\alpha_X = 1.0$. Space densities are given as
number of objects per $\rm Gpc^{3}$ and X-ray luminosity bin following
Padovani \& Giommi (\cite{emss}). The data from the EMSS are consistent
with those from the HRX-BL~Lac {\it complete sample} within the $1
\sigma$ error bars. The marginal differences can be due to systematic
errors for the calculated luminosities in the IPC band because of
differing spectral slopes, or resulting from differences in the
calibration of the IPC and the PSPC detectors. 

In the right panel of Figure \ref{fig:hbldiffcompare} we compare the
differential luminosity function of the {\em complete sample} with the
corresponding function for AGNs at z$<$0.5.  The AGN X-ray luminosity
function was taken from the ROSAC sample (``A ROSAT based Search for
AGN-Clusters'', Tesch \cite{tesch}). This AGN sample was constructed
similarly as the HRX-BL Lac sample and both samples match closely in
brightnesses and redshifts. The ROSAC-AGN sample contains 182
RASS-AGNs with $z < 0.5$ identified in an area of $363 \,{\rm deg}^2$
in the constellation of Ursa Major.  The AGN X-ray luminosities have
been corrected for the different X-ray band ($0.1 - 2.4 \keV$ instead
$0.5 - 2.0 \keV$) using the same spectral slopes used for the ROSAC
sample.

We find that the space density of BL Lacs in the luminosity range $44
< \log\,L_{\rm X} < 46$ is about 10\% of the space density of
AGNs. In case that all AGNs have jets and would be classified as 
BL Lacs when looking into their jet,
an jet opening angle of $\sim 50 \grad$ would follow. But as the
jet emission is expected to be beamed, the BL Lacs appear to be
brighter than they are. Following Urry \& Shaefer (\cite{urryshaefer})
the observed luminosity is $L_{obs} = \delta^p \, L_{emi}$ with $L_{emi}$
being the emitted luminosity,  and
\begin{equation}
\delta = \frac{1}{\gamma ( 1- \frac{v}{c_0} \cos \theta)}
\end{equation}
where $\gamma$ is the Lorentz factor of the jet emission. $p$ depends
on the spectral slope and the jet flow model. For the simple case of a
moving blob and continuous reacceleration (Lind \& Blandford
\cite{lind}) which applies e.g. for the model of a wide X-ray jet
(Celotti et al. \cite{celotti}) the exponent is $p = 3 + \alpha$,
where $\alpha \simeq 1$ is the spectral index. For a conical jet this
exponent is $p = 2 + \alpha$ (Urry \& Shaefer \cite {urryshaefer}).
Assuming an jet opening angle of $\theta \sim 30 \grad$ (Urry \&
Padovani \cite{urry}) and a Lorentz factor $\gamma \sim 5$ the
amplification factor is $\delta^p \simeq 50$ (for the conical jet) and  $\delta^p \simeq 200$ (for the wide X-ray jet). 
Correcting the
luminosities accordingly yields a fraction of $\le 0.1 \%$ BL Lacs among
all  AGNs.
A smaller opening angle and/or larger
Lorentz factor would lead to an even lower BL Lac fraction among the
AGN.

\section{Discussion}
The large number of objects included in the HRX-BL Lac sample
enabled us to study in detail a number of BL Lac properties in different
wavelength regions. In the following, we will discuss these properties
with special emphasis on their compatibility with the most recent
models trying to unify the HBL and LBL objects.

\subsection{Strength of the calcium break and luminosities}
We observed a clear anti-correlation between calcium break strength
and the luminosity in the radio, near infrared, optical, and X-rays
bands.  We explain this with a wide range of luminosities for the
non-thermal source, while the host galaxies seem to have approximately
constant luminosity. The higher the luminosity of the central source,
the more does the core outshine the hosting galaxy, leading to
decreasing break strength with increasing luminosity.

The identification of several HRX-BL Lac with $25 \% < \rm Ca-break
< 40 \%$ and the smooth extension of the correlation between break
strength and luminosities into this range, supports previous findings that BL Lacs can have $\rm Ca-break
> 25\%$ (March\~a et al. \cite{marcha}; Laurent-Muehleisen et al. \cite{RGB}; Rector et al. \cite{emssbllac}; Landt et al. \cite{landt}) in contrast to earlier suggestions (Stocke et al. \cite{oldcabreak}).

Landt et al. (\cite{landt}) studied in detail the dependency of the Calcium break strength on the luminosity of the 
blazars inside the DXRBS sample (Padovani \cite{padovanidxrbs}). They found the same correlation as described here for the HRX-BL Lac sample and show that the Calcium break values decrease with increasing jet power, and therefore with increasing luminosity. Based on this they conclude that the break value of BL Lac objects could be an indicator of the orientation. 
Nevertheless different
luminosities of the core component (i.e. of the jet) will also play a major role in the effect, and it seems to be difficult to disentangle the influence from different orientation and from different jet luminosity. Thus though we find for the HRX-BL Lac sample the same correlation as described by Landt et al. \cite{landt} we conclude here only that the break strength is an indicator of different apparent luminosity, either based on different orientation, or on different jet power, or on mixture of both effects.

\subsection{HBL/LBL: evolutionary dichotomy?}
Most of the results for the HRX-BL Lac {\em core sample} (Bade et
al. \cite{bade}) could be confirmed by the {\em complete sample}
presented here. The most significant discrepancy are the different
results of the $V_{\rm e}/V_{\rm a}$ test. We found negative evolution
using the {\em complete sample}, but we did not find differences in
evolution, if we divide the sample by $\alpha_{OX}$ (or peak
frequency). 

%
%
The results of Bade et al. (\cite{bade}) could have been arisen from
selection effects due to the ``patchy'' search area used.  A
Monte-Carlo simulation done on the HRX-BL Lac {\em complete sample}
shows however, that this is not the case. 
By randomly selecting a subsample of 17 BL Lac objects
(which is the number of objects for which Bade et al. found different
types of evolution) out of the HBLs of the {\em complete sample} there
is a chance of $< 1\%$ only to find a $\langle V_{\rm e}/V_{\rm a} \rangle
< 0.35$. 

%
%

Another reason for the different results could be related to the
different treatment of the radio properties. As radio detection was
not a selection criterium in Bade et al., no radio flux limit was
taken into account. Applying here the $V_{\rm e}/V_{\rm a}$-test to
the {\em complete sample} (cf. Sect. \ref{raum}), the accessible
volume $V_{\rm a}$ was determined for $\approx$10\% of the objects by
the radio limit.  The $V_{\rm e}/V_{\rm a}$ values are correspondingly
increased compared to the case where only X-ray flux limit is taken
into account, resulting in a less negative evolution. There remains the
fact that no BL Lac objects were found yet, with $f_{\rm R} < 2 \; \rm mJy$ 
radio counterparts and the question is still open whether
this is a selection effect or not. It could be that our decision to apply
the radio detection as selection criterium weakens the negative evolution
found in pure X-ray selected samples.

The result, that the evolution of BL Lac objects of the HBL and IBL
type is consistent with no evolution is in good agreement with other
recent studies. Neither REX or DXRBS, nor the HRX-BL Lac sample show a
difference for the more or less X-ray dominated BL Lacs. On the
contrary it seems that the evolution of the IBL might even be slightly
more positive than that of the HBL class.  This picture clearly
differs from the EMSS BL Lac result of $\langle V_{\rm e}/V_{\rm a}
\rangle = 0.36 \pm 0.05$, while the sendentary survey, presented by
Giommi, Menna \& Padovani \cite{sedentary} seems not be complete
enough up to now to draw a firm conclusion.  Caccianiga et
al. (\cite{rex3}) argue, that the REX might miss the negative
evolution of the HBL is not visible simply because the sample is not
deep enough, and this argumentation then would also apply for the
HRX-BL Lac sample, which X-ray flux limit is about two times higher
than that of the REX BL Lacs. Their simulation result in the
conclusion that even a completion of the REX survey might not lead to
a highly significant negative evolution ($2 \sigma$ for the simulated
sample).

Finally, the evolution found in the course of this work is
in good agreement with that of FR-I galaxies ($\langle
V_{\rm e}/V_{\rm a} \rangle = 0.40 \pm 0.06$) within the 3CR sample (Laing et
al. 1984).
 This supports the assumption that the FR-I galaxies build
the parent population of BL Lac objects (see e.g. Padovani \& Urry
\cite{padourry}).

In contrast to HBLs, the LBLs show weak or positive evolution
($\langle V_{\rm e}/V_{\rm a} \rangle = 0.61 \pm 0.05$) as shown for the 1Jy
sample by Rector \& Stocke (\cite{rector}). Following the sequence of blazars, also FSRQs exhibit significant positive evolution ($\langle V_{\rm e}/V_{\rm a} \rangle = 0.58 \pm 0.03$ for the 119 FSRQs in the DXRBS sample; Padovani \cite{padovanidxrbs}). Also FR-II radio galaxies and ``normal'' quasars seem to be more numerous and/or luminous
at cosmological distances than in the neighborhood, leaving the
question for the reasons of the HBL/LBL evolutionary dichotomy of
relevance also in future.

\subsection{A unifying model for LBL and HBL}
The different evolutionary behaviour of HBLs and LBLs is a challenge
for all theories to unify both BL Lac types into one class.
However, the existence of transition objects and the numerous similar
properties of LBLs and HBLs make it plausible that both classes
belong to the same parent population. 

As described by  B\"ottcher \&
Dermer (\cite{boettcher}) one way to unify both classes 
would be a transformation of LBLs into HBLs as
the BL Lac objects grow older. In this model, BL Lac objects start
as LBLs with jets of high energy densities. Strong cooling
limits the electron energies leading to cutoff frequencies for
the synchrotron component at optical wavelengths and for the IC
component in the GeV energy range. As shown by Beckmann et al. 
(\cite{saxpap}), this results in steep X-ray spectra with strong
curvature. The core outshines the host galaxy leading
to a low calcium break value (Landt et al. \cite{landt}) as seen also for the HRX-BL Lac sample (cf. Fig. \ref{fig:hblbreaklum}).

When by the time the source of the jet gets less powerful the energy
density within the jet decreases 
(Tavecchio et al. \cite{tavecchio}). The cooling efficiency decreases as
well resulting in higher cutoff frequencies for HBLs. 
The shift of the cutoff frequencies to higher energies is
therefore accompanied by decreasing bolometric luminosities,
which is evident from the decrease of the luminosities
in the radio, near IR and optical bands.
Due to the increasing peak frequencies of the synchrotron
branch more energy is released in the X-ray band and the X-ray luminosity 
increases quite in contrast to the luminosities at shorter frequencies
(cf. Table \ref{hblpeaklumtab}). The X-ray spectra are correspondingly
flatter and less curved than in the LBL state (Beckmann \& Wolter
\cite{bewo}). 

Objects which do not fit into this scenario are doubtlessly the
extremely luminous HBLs, like \object{1ES 1517+656} (Beckmann et
al. \cite{1517}).  The scenario presented here assumes the HBLs to be
on average less luminous than the LBLs. Apart from the exceptionally
high X-ray luminosity, this object also shows an optical luminosity
typical for a Flat Spectrum Radio Quasar (FSRQ). Padovani
\cite{padovanidxrbs} argues that those high state BL Lacs with high
peak frequency might belong to the high energy peaked FSRQ class
(HFSRQ), flat-spectrum radio quasars with synchrotron peak in the
UV/X-ray band. In this case 1ES 1517+656 should show strong emission
lines, e.g. stron $H_{\alpha}$ and $H_{\beta}$ which would be located
in the near infrared for this high-redshift BL Lac ($z = 0.7$) and
would have been missed by previous observations.

The HRX-BL~Lac sample could be the basis to study the extreme end
of the HBL population, the ultra high frequency peaked BL Lac objects
(UHBL). Sambruna et al. (\cite{sambruna}) argued that objects with
cutoff frequencies higher than $10^{18} \Hz$ would be detected only in
hard X-ray surveys but should be faint at lower frequencies, which
would make their discovery difficult.

Nevertheless HBLs have already been detected at TeV energies, as e.g.
1ES 1426+428 (Aharonian et al. \cite{1426a}; Horan et al. \cite{1426})
and 1ES1959+650 (Horns \& Konopelko \cite{1959}). Recently Costamante
\& Ghisellini (\cite{tevcandidates}) showed that it is possible to
select candidates for TeV BL Lacs on the basis of the knowledge of the
SED, i.e. strong X-ray flux and a sufficiently strong
radio-through-optical flux, which results in high peak frequencies of
the synchrotron branch.

Also 13 HBLs within the
HRX-BL~Lac sample show peak frequencies $\nu_{peak} > 10^{18} \Hz$
from the parabolic fit to the synchrotron branch and three objects
even $\nu_{peak} > 10^{19} \Hz$. 1RXS J121158.1+224236 might even be a UHBL
with a peak frequency of the synchrotron branch at $\nu_{peak} \simeq 10^{22} \Hz$ . To confirm the high peak frequencies, for this extreme source, 
observations with the BeppoSAX satellite have been performed and
results will be presented in a forthcoming paper. 
Investigations in the
gamma region ($\sim 1 \, {\rm MeV}$) are needed to decide whether
these energies are dominated by the synchrotron emission or if already
the inverse Compton branch is rising. The SPI spectrograph on-board
the INTEGRAL mission (see e.g. Winkler \& Hermsen \cite{INTEGRAL}),
which has been successfully launched in October 2002, will allow to do
spectroscopy in this energy region ($20 \keV - 8 \, {\rm MeV}$). 

\begin{acknowledgements}
We would like to thank the anonymous referee for the valuable
suggestions which helped us to improve the paper.
This research has made use of the \linebreak NASA/IPAC Extragalactic
Database (NED) which is operated by the Jet Propulsion Laboratory,
California Institute of Technology under contract with the National
Aeronautics and Space Administration.
We acknowledge support by the Deutsche Forschungsgemeinschaft through grants Re 353/39-1 and En 176/23-1.
\end{acknowledgements}

\newpage
\onecolumn
\section{Appendix: Tables to the HRX-BL Lac sample}
\ \\
\tablefirsthead{
    \multicolumn{1}{c}{$\alpha$ } 
  & \multicolumn{1}{c}{$\delta$} 
  & \multicolumn{1}{c}{Name}
  & \multicolumn{1}{c}{Redshift$^a$}
  & \multicolumn{1}{c}{Classification$^a$}\\
\noalign{\smallskip}\hline\noalign{\smallskip}
}
\tablehead{\multicolumn{3}{l}{\footnotesize \em
    The objects of the BSC/NVSS correlation}\\
    \multicolumn{1}{c}{$\alpha$ } 
  & \multicolumn{1}{c}{$\delta$} 
  & \multicolumn{1}{c}{Name}
  & \multicolumn{1}{c}{Redshift$^a$}
  & \multicolumn{1}{c}{Classification$^a$}\\
\noalign{\smallskip}\hline\noalign{\smallskip}
}
\tabletail{}
\tablelasttail{\multicolumn{3}{r}{\footnotesize \em (continued on the next
    page)}\\
\multicolumn{5}{l}{$^a$ new entries first presented here marked with $^*$,
  confirmed identifications/redshifts: $^C$}\\
}
\tablecaption{Objects from the RASS-BSC/NVSS correlation. An X-ray count rate
  limit $hcps \ge 0.09$ was applied.}
\begin{center}
\begin{supertabular}{lllll}
07 04 27.0 & 63 18 56 & KUG 0659+633          & 0.095 &  G\\
07 07 13.5 & 64 35 58 & VII Zw 118            & 0.080 &  Sy1\\
07 09 07.6 & 48 36 57 & NGC 2329              & 0.019 &  G S0-:\\
07 10 30.0 & 59 08 08 & 87GB 070609.2+591323  & 0.125 &  BL Lac\\
07 11 48.0 & 32 19 02 & PGC 020369            & 0.067 &  Sy2\\
07 13 39.7 & 38 20 43 & IRAS F07102+3825      & 0.123 &  NLSy1\\
07 18 00.7 & 44 05 27 & IRAS F07144+4410      & 0.061 &  Sy1 \\
07 21 53.2 & 71 20 31 & 0716+714              &       &  BL Lac\\
07 22 17.4 & 30 30 52 & HS 0719+3036          & 0.100 &  Sy1.5\\
07 32 21.5 & 31 37 50 & 1RXS J073221.5+313750 & 0.170 &  G \\
07 36 24.8 & 39 26 08 & FBS 0732+396          & 0.118 &  Sy1 \\
07 40 58.5 & 55 25 33 & CGCG 262-019          & 0.034 &  GGroup \\
07 41 44.8 & 74 14 45 & ZwCl 0735.7+7421      & 0.216 &  GClstr \\
07 42 32.9 & 49 48 30 & UGC 03973             & 0.022 &  Sy1.2\\
07 42 50.3 & 61 09 31 & 87GB 073825.7+611711  &       &  star (K0)\\
07 44 05.6 & 74 33 56 & MS 0737.9+7441        & 0.315 &  BL Lac\\
07 45 41.2 & 31 42 50 & [HB89] 0742+318       & 0.461 &  Sy1\\
07 47 29.4 & 60 56 01 & UGC 04013             & 0.029 &  Sy1.2\\
07 49 06.2 & 45 10 40 & B3 0745+453           & 0.190 &  Sy1\\
07 49 29.4 & 74 51 42 & 87GB[BWE91] 0743+7458 & 0.607$^*$ &  BL Lac$^*$\\
07 51 22.1 & 55 11 56 & 1RXS J075122.1+551156 & 0.302$^*$ &  Sy1$^*$\\
07 52 43.6 & 45 56 53 & NPM1G +46.0092        & 0.060 &  Sy1.9 \\
08 01 32.3 & 47 36 19 & 87GB 075755.8+474440  & 0.158 &  Sy1\\
08 01 47.8 & 56 33 16 & NGC 2488              & 0.029 &  G,  S0-:\\
08 05 25.8 & 75 34 24 & WN B0759.1+7542       & 0.121 &  BL Lac\\
08 06 25.5 & 59 31 05 & 87GB 080212.8+593933  &       &  BL Lac\\
08 09 38.5 & 34 55 44 & MG2 J080937+3455      & 0.082 &  BL Lac$^C$\\
08 09 49.2 & 52 18 55 & 87GB 080601.8+522753  & 0.138 &  BL Lac\\
08 10 59.0 & 76 02 45 & PG 0804+761           & 0.100 &  Sy1\\
08 15 17.8 & 46 04 29 & KUG 0811+462          & 0.041 &  Sy1.5\\
08 19 26.6 & 63 37 41 & KOS NP6 038           & 0.118$^C$ &  G, E$^C$\\
08 19 29.5 & 70 42 21 & 1RXS J081929.5+704221 & 0.001 &  HolmbergII\\
08 22 09.5 & 47 06 01 & RGB J0822+470         & 0.127 &  Sy1.5 \\
08 32 25.1 & 37 07 37 & FIRST J083225.3+370737 & 0.091 &  Sy1.2\\
08 32 51.9 & 33 00 11 & 1RXS J083251.9+330011 & 0.671 &  BL Lac\\
08 36 58.3 & 44 26 13 & [HB89] 0833+446       & 0.255 &  QSO\\
08 38 11.0 & 24 53 36 & NGC 2622              & 0.029 &  Sy1.8\\
08 41 25.1 & 70 53 43 & [HB89] 0836+710       & 2.172 &  blazar\\
08 42 03.4 & 40 18 30 & KUV 08388+4029        & 0.152 &  Sy1\\
08 42 55.9 & 29 27 52 & ZwCl 0839.9+2937      & 0.194 &  GClstr\\
08 59 16.5 & 83 44 50 & 1RXS J085916.5+834450 & 0.327$^*$ &  BL Lac$^*$\\
08 59 30.1 & 74 55 10 & 1RXS J085930.1+745510 & 0.276 &  Sy1 \\
09 09 53.9 & 31 05 58 & MG2 J090953+3104      & 0.274$^*$ &  BL Lac$^*$\\
09 13 24.6 & 81 33 18 & 1RXS J091324.6+813318 & 0.639$^*$ &  BL Lac$^*$\\
09 15 52.2 & 29 33 35 & B2 0912+29 / TON 0396 &       &  BL Lac$^C$\\
09 16 51.8 & 52 38 29 & 87GB 091315.6+525108  & 0.190 &  BL Lac\\
09 23 43.0 & 22 54 37 & CGCG 121-075          & 0.032 &  Sy1\\
09 25 12.3 & 52 17 17 & MRK 0110              & 0.035 &  Sy1\\
09 27 02.8 & 39 02 21 & [HB89] 0923+392       & 0.695 &  Sy1\\
09 28 04.2 & 74 47 14 & 87GB 092308.0+745942  & 0.638 &  BL Lac\\
09 30 37.1 & 49 50 28 & 1ES 0927+500          & 0.186 &  BL Lac\\
09 33 46.5 & 62 49 43 & 1RXS J093346.5+624943 &       &  star G5\\
09 35 27.4 & 26 17 14 & 1RXS J093527.4+261714 & 0.122 & Sy1\\
09 47 13.2 & 76 23 17 & RBS 0797              & 0.354 &  LINER\\
09 52 25.8 & 75 02 16 & 1RXS J095225.8+750216 & 0.178$^*$ &  BL Lac$^*$\\
09 55 34.7 & 69 03 38 & MESSIER 081           &-0.00011 & LINER/Sy1.8\\
09 55 50.4 & 69 40 52 & MESSIER 082           & 0.00068 & Starburst G\\
09 59 29.8 & 21 23 40 & 87GB 095643.1+213755  & 0.365$^C$ &  BL Lac$^C$\\
10 00 28.9 & 44 09 10 & 1RXS J100028.9+440910 & 0.154 &  GClstr\\
10 02 35.9 & 32 42 19 & NGC 3099              & 0.051 &  G\\
10 05 42.2 & 43 32 44 & IRAS  10026+4347      & 0.178$^C$ &  Sy1$^C$\\
10 06 39.1 & 25 54 51 & PGC 029375            & 0.116 &  GClstr\\
10 08 11.5 & 47 05 26 & 1RXS J100811.5+470526 & 0.343 &  BL Lac$^C$\\
10 09 16.7 & 71 10 40 & 87GB 100504.8+712548  & 0.193 &  GClstr\\
10 10 27.9 & 41 32 42 & [HB89] 1007+417       & 0.612 &  QSO\\
10 12 44.4 & 42 29 59 & B3 1009+427           & 0.376$^C$ &  BL Lac$^C$\\
10 15 04.3 & 49 26 04 & [HB89] 1011+496       & 0.200 &  BL Lac\\
10 16 16.4 & 41 08 17 & FIRST J101616.8+410812& 0.281 &  BL Lac\\
10 16 55.5 & 73 23 59 & NGC 3147              & 0.009 &  Sy2\\
10 17 18.0 & 29 14 39 & IRAS F10144+2929      & 0.048 &  Sy1\\
10 19 00.8 & 37 52 50 & FIRST J101900.4+375240& 0.133 &  Sy1\\
10 19 12.1 & 63 58 03 & MRK 141               & 0.042 &  Sy1.5\\
10 22 12.5 & 51 24 06 & MS 1019.0+5139        & 0.141 &  BL Lac\\
10 22 28.9 & 50 06 30 & ABELL 0980:[CAE99]    & 0.158 &  G\\
10 30 58.8 & 31 03 06 & [HB89] 1028+313       & 0.178 &  Sy1\\
10 31 05.7 & 82 33 27 & 1RXS J103105.7+823327 &       &  star F2V\\
10 31 18.6 & 50 53 41 & 1ES 1028+511          & 0.361 &  BL Lac$^C$\\
10 32 14.3 & 40 16 07 & [BBN91] 102920+403136 & 0.078 &  GClstr\\
10 34 23.1 & 73 45 25 & NGC 3252              & 0.004 &  G, SBd?sp\\
10 34 38.7 & 39 38 34 & KUG 1031+398          & 0.042 &  Sy1\\
10 34 59.5 & 30 41 39 & Abel 1045             & 0.137 &  GClstr\\
10 38 46.7 & 53 30 02 & SN 1991N              & 0.003 &  SNR in NGC3310\\
10 40 43.7 & 39 57 06 & IRAS F10378+4012      & 0.139 &  Sy2\\
10 44 27.6 & 27 18 13 & 1RXS J104427.6+271813 &       &  G\\
10 44 39.4 & 38 45 42 & CGCG 212-045          & 0.036 &  Sy1.5\\
10 45 20.5 & 45 34 04 & 1RXS J104520.5+453404 &       &  star B8V\\
10 51 25.1 & 39 43 29 & FIRST J105125.3+394325& 0.498$^*$ &  BL Lac$^*$\\
10 55 44.0 & 60 28 10 & 1RXS J105544.0+602810 &       &  var. K0 star\\
10 57 23.5 & 23 03 17 & 1RXS J105723.5+230317 & 0.378$^C$ &  BL Lac$^C$\\
10 58 25.9 & 56 47 16 & RX J1058+5647         &       &prob. GClstr$^*$\\
10 58 37.5 & 56 28 16 & 87GB 105536.5+564424  & 0.144 &  BL Lac\\
11 00 21.3 & 40 19 33 & FIRST J110021.0+401927& 0.225$^*$ &  BL Lac$^C$\\
11 04 12.4 & 76 58 59 & PG 1100+772           & 0.312 &  QSO
(Opt.var.)\\         
11 04 27.1 & 38 12 32 & MRK 421               & 0.030 &  BL Lac\\
11 06 43.5 & 72 34 07 & NGC 3516              & 0.009 &  Sy1.5\\
11 11 31.2 & 34 52 12 & FIRST J111130.8+345203& 0.212 &  BL Lac\\
11 11 37.2 & 40 50 31 & ABELL 1190:[SBM98]    & 0.079 &  GClstr\\
11 14 22.6 & 58 23 18 & 8C 1111+586           & 0.206 &  GClstr\\
11 17 06.3 & 20 14 10 & 87GB 111429.0+203022  & 0.137$^C$ &  BL Lac$^C$\\
11 18 03.6 & 45 06 57 & LEDA 139560           & 0.106 &  Sy1\\
11 19 08.1 & 21 19 15 & PG 1116+215           & 0.176 &  Sy1\\
11 20 47.5 & 42 12 17 & 1ES 1118+424          & 0.124 &  BL Lac\\
11 21 09.9 & 53 51 25 & 1RXS J112109.9+535125 & 0.103 &  Sy1 \\
11 23 49.2 & 72 30 02 & 1RXS J112349.2+723002 &       &  BL Lac$^C$\\
11 23 57.4 & 21 29 14 & [CWH99] 112356.7+2129 & 0.199 &  GClstr \\
11 31 08.9 & 31 14 09 & TON 0580              & 0.289 &  QSO\\
11 31 21.4 & 33 34 47 & 1RXS J113121.4+333447 & 0.220$^C$ &  G$^C$\\
11 32 22.4 & 55 58 28 & 1RXS J113222.4+555828 & 0.051 &  GClstr \\
11 36 26.6 & 70 09 32 & MRK 180               & 0.046 &  BL Lac\\
11 36 29.4 & 21 35 53 & MRK 739b              & 0.030 &  Sy1\\
11 36 30.9 & 67 37 08 & HS 1133+6753          & 0.135 &  BL Lac\\
11 41 16.2 & 21 56 25 & PG 1138+222           & 0.063 &  Sy1\\
11 44 29.9 & 36 53 14 & KUG 1141+371          & 0.040 &  Sy1\\
11 45 09.3 & 30 47 25 & [HB89] 1142+310       & 0.060 &  Sy1.5\\
11 47 54.9 & 22 05 48 & 1RXS J114754.9+220548 & 0.276$^*$ &  BL Lac$^*$\\
11 49 30.4 & 24 39 28 & RGB J1149+246         & 0.402$^*$ &  BL Lac$^C$\\
11 53 24.4 & 49 31 09 & [HB89] 1150+497       & 0.334 &  QSO \\
11 55 18.9 & 23 24 32 & MCG +04-28-097        & 0.143 &  G\\
11 57 56.1 & 55 27 17 & NGC 3998              & 0.003 &  Sy1;LINER\\
12 03 08.9 & 44 31 55 & NGC 4051              & 0.002 &  Sy1.5\\
12 03 43.3 & 28 36 02 & 1RXS J120343.3+283602 & 0.373 &  Sy1\\
12 05 11.7 & 39 20 43 & 1RXS J120511.7+392043 & 0.037 &  GClstr \\
12 09 46.0 & 32 17 03 & FIRST J120945.2+321700& 0.145 &  Sy1\\
12 11 58.1 & 22 42 36 & 1RXS J121158.1+224236 & 0.455$^*$ &  BL Lac$^*$\\
12 13 45.2 & 36 37 55 & NGC 4190              & 0.001 &  G \\
12 15 06.7 & 33 11 30 & NGC 4203              & 0.004 &  LINER\\
12 17 52.1 & 30 07 05 & ON 325 / B2 1215+30   & 0.130 &  BL Lac\\
12 18 27.0 & 29 48 53 & NGC 4253              & 0.013 &  Sy1.5\\
12 21 21.7 & 30 10 41 & FBQS J1221+3010       & 0.182 &  BL Lac\\
12 24 54.9 & 21 22 52 & PG 1222+216           & 0.435 &  blazar\\
12 25 12.5 & 32 13 54 & MAPS-NGP O-267-076115 & 0.059 &  GClstr\\
12 26 23.9 & 32 44 31 & NGC 4395:[R97] 12     & 0.242$^C$ &  Sy1$^C$\\
12 30 14.2 & 25 18 05 & MG2 J123013+2517      & 0.135 &  BL Lac$^C$\\
12 31 32.5 & 64 14 20 & [HB89] 1229+645       & 0.164$^*$ &  BL Lac$^*$\\
12 32 03.6 & 20 09 30 & MRK 771               & 0.063 &  Sy1\\
12 36 51.1 & 45 39 07 & CGCG 244-033          & 0.030 &  Sy1.5\\
12 36 58.8 & 63 11 11 & ABELL 1576:[HHP90]    & 0.302 &  G\\
12 37 05.6 & 30 20 02 & FIRST J123705.5+302005& 0.700 &  BL Lac$^C$\\
12 37 39.2 & 62 58 43 & [HB89] 1235+632       & 0.297 &  BL Lac\\
12 38 08.3 & 53 26 04 & 87GB 123550.3+534219  & 0.347$^C$ &  Sy1$^C$\\
12 39 23.1 & 41 32 45 & FIRST J123922.7+413251&       &  BL Lac$^*$\\
12 41 41.2 & 34 40 32 & FIRST J124141.3+344031&       &  BL Lac$^C$\\
12 41 44.4 & 35 03 53 & NGC 4619              & 0.023 &  Sy1\\
12 42 11.3 & 33 17 03 & WAS 61                & 0.044 &  Sy1\\
12 43 12.5 & 36 27 43 & TON 0116              &       &  BL Lac$^C$ \\
12 47 01.3 & 44 23 25 & RGB J1247+443         &       &  AGN \\
12 48 18.9 & 58 20 31 & PG 1246+586           &       &  BL Lac$^C$ \\
12 50 52.5 & 41 07 13 & MESSIER 94            & 0.001 &  LINER \\
12 53 01.0 & 38 26 29 & FIRST J125300.9+382625& 0.372$^C$&  BL Lac$^C$ \\
12 57 31.7 & 24 12 46 & 1ES 1255+244          & 0.141 &  BL Lac \\
13 02 55.6 & 50 56 21 & 1RXS J130255.6+505621 & 0.688 &  BL Lac$^C$ \\
13 05 52.6 & 30 54 06 & ABELL 1677            & 0.183 &  GClstr \\
13 13 27.2 & 36 35 42 & NGC 5033              & 0.003 &  Sy1.9 \\
13 19 57.2 & 52 35 33 & 1RXS J131957.2+523533 & 0.092 &  Sy \\
13 20 16.3 & 33 08 29 & 1RXS J132016.3+330829 & 0.036 &  GClstr \\
13 22 48.5 & 54 55 27 & 1RXS J132248.5+545527 & 0.064 &  Sy1 \\
13 24 00.2 & 57 39 18 & 87GB 132204.6+575429  & 0.115 &  BL Lac\\
13 25 49.2 & 59 19 37 & ABELL 1744:[HHP90]    & 0.151 &  GClstr  \\
13 26 15.0 & 29 33 32 & 87GB 132354.7+294853  & 0.431 &  BL Lac \\
13 34 47.5 & 37 11 00 & BH CVn                &       &  star
(F2IVRSCVn D)\\
13 35 08.2 & 20 46 41 & 1RXS J133508.2+204641 &       &  var. K0 star\\
13 37 18.8 & 24 23 07 & [HB89] 1334+246       & 0.108 &  Sy    D\\
13 40 29.9 & 44 10 07 & 87GB 133822.3+442514  & 0.548$^*$ &  BL Lac$^C$\\
13 41 04.8 & 39 59 42 & B3 1338+402           & 0.163 &  BL Lac \\
13 41 52.6 & 26 22 30 & 1RXS J134152.6+262230 & 0.075 &  GClstr$^*$ \\
13 45 45.1 & 53 33 01 & 87GB 134352.4+534755  & 0.135$^*$ &  Sy1$^C$ \\
13 48 52.6 & 26 35 41 & ABELL 1795:[MK91]     & 0.062 &  GClstr \\
13 53 28.2 & 56 01 02 & 1RXS J135328.2+560102 & 0.370 &  BL Lac \\
13 54 20.2 & 32 55 47 & UGC 08829             & 0.026 &  Sy1 \\
13 55 15.9 & 56 12 44 & 1RXS J135515.9+561244 & 0.122 &  Sy1 \\
13 55 53.3 & 38 34 28 & MRK 464               & 0.051 &  Sy1.5 \\
14 04 50.2 & 65 54 34 & 1RXS J140450.2+655434 & 0.364 &  BL Lac$^C$\\
14 06 22.2 & 22 23 50 & PG 1404+226           & 0.098 &  Sy \\
14 10 31.6 & 61 00 20 & 1RXS J141031.6+610021 & 0.384 &  BL Lac$^C$\\
14 13 42.6 & 43 39 38 & MAPS-NGP O-221-004710 & 0.089 &  G  \\
14 13 58.3 & 76 44 56 & 1RXS J141358.3+764456 & 0.068$^*$ &  Sy2$^*$\\
14 17 56.8 & 25 43 29 & 1E 1415+259           & 0.237 &  BL Lac \\
14 17 59.6 & 25 08 18 & SN 1984Z              & 0.017 &  SNR in
NGC5548 \\           
14 21 36.4 & 49 33 05 & MCG +08-26-021        & 0.072 &  GClstr \\
14 21 39.7 & 37 17 43 & 1RXS J142139.7+371743 & 0.160 &  GClstr \\
14 22 39.1 & 58 01 59 & RGB J1422+580         & 0.638 &  BL Lac$^C$ \\
14 23 13.4 & 50 55 37 & 87GB 142127.2+510856  & 0.274 &  Sy1 \\
14 23 53.6 & 40 15 33 & 1RXS J142353.6+401533 & 0.082 &  GClstr \\
14 23 56.0 & 26 26 30 & 1RXS J142356.0+262630 &       &Prob. GClstr$^*$\\
14 26 01.3 & 37 49 36 & ABELL 1914            & 0.171 &  GClstr\\
14 27 00.5 & 23 48 03 & PG 1424+240           &       &  BL Lac \\
14 28 32.6 & 42 40 28 & 1ES 1426+428          & 0.129 &  BL Lac\\
14 31 04.8 & 28 17 16 & MRK 684               & 0.046 &  Sy1 \\
14 31 06.2 & 25 38 15 & 1RXS J143106.2+253815 & 0.096 &  GClstr \\
14 32 36.0 & 31 38 55 & 1RXS J143236.0+313855 & 0.132 &  GClstr \\
14 39 17.7 & 39 32 48 & PG 1437+398           &       &  BL Lac$^C$ \\
14 42 07.7 & 35 26 32 & MRK 478               & 0.079 &  Sy1 \\
14 42 18.9 & 22 18 20 & UGC 09480             & 0.097 &  GClstr \\
14 43 02.8 & 52 01 41 & 3C 303                & 0.141 &  G \\
14 44 33.9 & 63 36 04 & MS 1443.5+6349        & 0.299 &  BL Lac \\
14 48 01.0 & 36 08 33 & [WB92] 1446+3620      &       &  BL Lac \\
14 49 32.3 & 27 46 30 & RBS 1434              & 0.228$^*$ &  BL Lac$^*$ \\
14 51 08.5 & 27 09 33 & PG 1448+273           & 0.065 &  Sy1 \\
14 56 03.4 & 50 48 24 & 1RXS J145603.4+504824 & 0.480 &  BL Lac \\
14 57 15.4 & 22 20 26 & MS 1455.0+2232        & 0.258 &  GClstr \\
14 58 27.3 & 48 32 50 & 1RXS J145827.3+483250 & 0.539 &  BL Lac \\
15 00 20.7 & 21 22 14 & LEDA 140447           & 0.153 &  G  \\
15 01 01.7 & 22 38 12 & MS 1458.8+2249        & 0.235 &  BL Lac \\
15 04 13.1 & 68 56 10 & [HB89] 1503+691       & 0.318 &  Sy1 \\
15 07 44.6 & 51 27 10 & MRK 845               & 0.046 &  Sy1 \\
15 08 42.2 & 27 09 10 & RBS 1467              & 0.270$^*$ &  BL Lac$^C$\\
15 10 40.8 & 33 35 15 & 1RXS J151040.8+333515 & 0.116$^C$ &  BL Lac$^C$ \\
15 14 43.1 & 36 50 59 & [HB89] 1512+370       & 0.371 &  QSO/Sy1? \\
15 17 47.3 & 65 25 23 & 1517+656              & 0.702$^C$ &  BL Lac$^C$ \\
15 21 53.0 & 20 58 30 & 1RXS J152153.0+205830 &       &  star M9 \\
15 23 46.0 & 63 39 30 & 4C +63.22             & 0.204 &  G  \\
15 29 07.5 & 56 16 05 & IRAS F15279+5626      & 0.099 &  Sy1 \\
15 32 02.3 & 30 16 31 & 87GB 152959.0+302636  & 0.064 &  BL Lac \\
15 32 53.7 & 30 21 03 & RBS 1509              & 0.361 &
LINER/GClstr \\      
15 33 24.9 & 34 16 40 & 87GB 153121.5+342710  &       &  BL Lac \\
15 35 01.1 & 53 20 42 & 1ES 1533+535          & 0.890 &  BL Lac$^C$ \\
15 35 52.0 & 57 54 04 & MRK 290               & 0.030 &  Sy1 \\
15 39 50.3 & 30 43 05 & 1RXS J153950.3+304305 & 0.097 &  GClstr \\
15 40 16.4 & 81 55 05 & 1ES 1544+820          &       &  BL Lac$^C$ \\
15 47 44.2 & 20 51 56 & 3C 323.1              & 0.264 &  QSO \\
15 54 24.3 & 20 11 16 & MS 1552.1+2020        & 0.222 &  BL Lac \\
15 58 18.7 & 25 51 18 & MRK  864              & 0.072 &  Sy2 \\
15 59 09.5 & 35 01 45 & UGC 10120             & 0.031 &  Sy1 \\
\end{supertabular}
\label{allcorr}
\end{center}

\tablefirsthead{
    \multicolumn{1}{c}{Name$^a$}
  & \multicolumn{1}{c}{$\alpha$} 
  & \multicolumn{1}{c}{$\delta$} 
  & \multicolumn{1}{c}{$z^b$}
  & \multicolumn{1}{c}{$f_{\rm X}^c$}
  & \multicolumn{1}{c}{$f_R^d$}
  & \multicolumn{1}{c}{B mag$^e$}
  & \multicolumn{1}{c}{K mag}
  & \multicolumn{1}{c}{Ca break}\\ 
\noalign{\smallskip}\hline\noalign{\smallskip}
}
\tablehead{\multicolumn{3}{l}{\footnotesize \em
        The HRX-BL Lac sample}\\
    \multicolumn{1}{c}{Name$^a$}
  & \multicolumn{1}{c}{$\alpha$ } 
  & \multicolumn{1}{c}{$\delta$} 
  & \multicolumn{1}{c}{$z^b$}
  & \multicolumn{1}{c}{$f_{\rm X}^c$}
  & \multicolumn{1}{c}{$f_R^d$}
  & \multicolumn{1}{c}{B mag$^e$}
  & \multicolumn{1}{c}{K mag}
  & \multicolumn{1}{c}{Ca break}\\ 
\noalign{\smallskip}\hline\noalign{\smallskip}
}
\tablelasttail{\multicolumn{9}{l}{$^a$ objects not included in the
    {\it complete sample} marked with an asterisk, new identifications
    marked with {\em new}}\\
\multicolumn{3}{l}{$^b$ uncertain redshifts are marked with an ``?''}\\
\multicolumn{4}{l}{$^c$ ROSAT PSPC (0.5 - 2.0 keV) flux in $10^{-12} \ecs$}\\
\multicolumn{3}{l}{$^d$ radio flux at $1.4 \GHz$ in mJy}\\         
\multicolumn{7}{l}{$^e$ B magnitudes determined with the Calar Alto
  1.23m telescope are marked with $^1$}\\        
}       
\tabletail{}
\tablecaption{The HRX-BL Lac sample}
\begin{center}
\begin{supertabular}{llllrrllc}
\object{1RXS J071030.0+590808}    & 07 10 30.1 & +59 08 20 & 0.125 & 10.15& 159.2 & 18.4 & &\\
\object{1RXS J071218.9+571948}$^*$& 07 12 18.9 & +57 19 48 & 0.095 & 1.01 &
7.9 & 20.1 & & $34 \%$\\
\object{0716+714}                 & 07 21 53.5 & +71 20 36 &       & 1.26 &
727.2 & 15.5 & 10.4& \\
\object{MS 0737.9+7441}           & 07 44 05.1 & +74 33 58 & 0.0311& 4.46 &
23.3 & 16.9 & 14.6 &\\
\object{1RXS J074929.4+745142}    & 07 49 29.7 & +74 51 45 & 0.605 & 2.80 &
44.8 & 18.9 & 15.2 & $2 \%$\\
\object{1RXS J080323.1+481622}$^*new$& 08 03 22.9 & +48 16 19 & 0.503 & 0.92 &
12.6 & 18.8 & & $5 \%$\\
\object{1RXS J080525.8+753424}    & 08 05 26.9 & +75 34 25 & 0.121 & 2.59 &
52.6 & 18.1 & 13.7 & \\
\object{1RXS J080625.5+593105}    & 08 06 25.9 & +59 31 06 &       & 2.06 &  60.9 & 17.9 & &\\
\object{1RXS J080938.5+345544}    & 08 09 38.5 & +34 55 37 & 0.082 & 2.62 &
223.4 & 17.0 & 14.5 & \\
\object{1RXS J080949.2+521855}    & 08 09 49.0 & +52 18 56 & 0.138 & 4.71 & 182.8 & 15.6 & &\\
\object{1RXS J081624.6+573910}$^*new$& 08 16 22.7 & +57 39 09 &       & 0.94 &
100.0 & 18.8 & & $5 \%$\\
\object{1RXS J083251.9+330011}    & 08 32 52.0 & +33 00 11 & 0.671 & 1.25 &
6.6 & 20.7 & & \\
\object{1RXS J083357.5+472658}$^*new$& 08 33 57.1 & +47 26 51 & 0.496 & 0.82 &
11.6 & 19.7 & & $14 \%$\\
\object{1RXS J085407.7+440840}$^*new$& 08 54 09.8 & +44 08 31 &       & 0.76 &
79.8 & 18.5 & 14.3 &\\
\object{1RXS J085451.0+621843}$^*new$& 08 54 50.5 & +62 18 50 & 0.267? & 0.83
& 387.6 & 19.0 & & \\
\object{1RXS J085916.5+834450}    & 08 59 10.1 & +83 45 04 & 0.327 & 1.23 &
10.2 & 19.7$^1$ & & $9 \%$\\
\object{1RXS J090315.3+405609}$^*$& 09 03 14.7 & +40 56 01 & 0.190 & 0.93 &
38.2 & 19.3 & & $25 \%$\\
\object{B2 0906+31}               & 09 09 53.3 & +31 06 02 & 0.274 & 2.18 &
195.5 & 18.3 & 14.1 & $3 \%$\\
\object{1RXS J091324.6+813318}    & 09 13 20.4 & +81 33 06 & 0.639 & 2.28 &
4.9 & 20.7$^1$ & & $5 \%$\\
\object{B2 0912+29}               & 09 15 52.2 & +29 33 20 &       & 3.36 &
342.0 & 16.3$^1$ & 12.7 &\\
\object{1RXS J091651.8+523829}    & 09 16 52.0 & +52 38 27 & 0.190 & 2.02 &
138.9 & 18.3$^1$ & 14.3 &\\
\object{1RXS J092401.1+053350}$^*new$& 09 24 01.1 & +05 33 50 &       & 1.34 &   7.5 & 19.6$^1$ & &\\
\object{1RXS J092804.2+744714}    & 09 28 03.0 & +74 47 19 & 0.638 & 1.13 &  85.8 & 20.8$^1$ & &\\
\object{1ES 0927+500}             & 09 30 37.6 & +49 50 24 & 0.186 &13.45 &  21.4 & 18.0 & &$26 \%$\\
\object{1RXS J093056.2+393332}$^*$& 09 30 56.9 & +39 33 37 & 0.641 & 0.80 &  13.0 & 19.5 & &$30 \%$\\
\object{1RXS J094020.7+614837}$^*new$& 09 40 22.5 & +61 48 25 & 0.212 & 0.85 &  12.8 & 18.4 & &$29 \%$\\
\object{1RXS J095214.4+393627}$^*new$& 09 52 14.0 & +39 36 08 &       & 0.65 &
3.0 & 19.8 & 15.7 &\\
\object{1RXS J095225.8+750216}    & 09 52 23.8 & +75 02 13 & 0.178 & 1.98 &
12.2 & 19.3$^1$ & 14.6 &\\
\object{1RXS J095410.8+491457}$^*$& 09 54 09.8 & +49 14 59 & 0.207 & 0.76 &   2.7 & 19.3 & &\\
\object{1RXS J095929.8+212340}    & 09 59 30.0 & +21 23 19 & 0.367 & 2.32 &
40.8 & 19.1$^1$ & 14.6 & $3 \%$\\
\object{1RXS J100656.9+345446}$^*new$& 10 06 56.3 & +34 54 44 & 0.612?& 0.66 &   6.6 & 18.7 & &$-14 \%$\\
\object{1RXS J100811.5+470526}    & 10 08 11.4 & +47 05 20 & 0.343 & 4.37 &   4.7 & 19.9$^1$ & &\\
\object{1RXS J101244.4+422959}    & 10 12 44.2 & +42 29 57 & 0.376 & 3.86 &
79.5 & 18.4$^1$ & 14.9 & $-2 \%$\\
\object{GB 1011+496}              & 10 15 04.0 & +49 25 59 & 0.200 & 6.78 &
378.1 & 16.5 & 13.1 & \\
\object{1RXS J101616.4+410817}    & 10 16 16.8 & +41 08 12 & 0.281 & 2.50 &
14.8 & 19.5$^1$ & 14.6 & \\
\object{MS 1019.0+5139}           & 10 22 11.3 & +51 24 15 & 0.141 & 3.35 &   5.1 & 18.0 & &\\
\object{1ES 1028+511}             & 10 31 18.6 & +50 53 34 & 0.361 & 18.02&  37.9 & 16.8 & 13.9\\
\object{1RXS J103744.9+571159}$^*$& 10 37 44.3 & +57 11 56 &       & 0.60 &  71.7 & 16.7 & &\\
\object{1RXS J104148.9+390133}$^*new$& 10 41 49.0 & +39 01 22 & 0.210 & 0.71 &
33.9 & 18.5 & 14.4 & $28 \%$\\
\object{1RXS J105125.1+394329}    & 10 51 25.4 & +39 43 26 & 0.498 & 1.67 &
10.8 & 19.0$^1$ & 15.4 & $-1 \%$\\
\object{1RXS J105607.0+025215}$^*$& 10 56 06.3 & +02 52 28 & 0.235 & 6.18 &   4.3 & 19.4$^1$ & &\\
\object{1RXS J105723.5+230317}    & 10 57 23.0 & +23 03 15 & 0.378 & 2.14 &   7.9 & 19.7$^1$ & &$13 \%$\\
\object{1RXS J105837.5+562816}    & 10 58 37.8 & +56 28 09 & 0.144 & 1.36 & 228.5 & 15.8 & &\\
\object{1RXS J110021.3+401933}    & 11 00 21.0 & +40 19 29 & 0.225?& 2.21 &
18.3& 18.6$^1$ & 15.2 & $-5 \%$\\
\object{MRK 421}                  & 11 04 27.3 & +38 12 32 & 0.030 &116.48& 768.5 & 13.3 & &\\
\object{1RXS J110748.2+150217}$^*new$& 11 07 48.2 & +15 02 17 &       & 2.17 &
43.5 & 18.4$^1$ & 14.8 & \\
\object{1RXS J111131.2+345212}    & 11 11 30.9 & +34 52 01 & 0.212 & 2.73 &   8.4 & 19.7$^1$ & &\\
\object{1RXS J111706.3+201410}    & 11 17 06.3 & +20 14 08 & 0.137 & 24.01&
103.1 & 16.0 & 12.4 & $-12 \%$\\
\object{1ES 1118+424}             & 11 20 48.1 & +42 12 12 & 0.124 & 4.63 &
24.1 & 18.0$^1$ & 14.7 &\\
\object{1RXS J112349.2+723002}    & 11 23 49.2 & +72 30 18 &       & 1.89 &
12.5 & 18.6 & 13.6 &\\
\object{MRK 180}                  & 11 36 26.6 & +70 09 25 & 0.046 & 19.73&
328.4 & 14.7 & 12.4 &\\
\object{HS 1133+6753}             & 11 36 30.3 & +67 37 05 & 0.135 & 11.40&
45.8 & 17.6 & 13.8 & \\
\object{1RXS J114754.9+220548}    & 11 47 54.9 & +22 05 34 & 0.276 & 1.39 &
4.1 & 21.0 & 15.0 &$33 \%$\\
\object{1RXS J114930.4+243928}    & 11 49 30.3 & +24 39 27 & 0.402 & 2.56 &  28.5 & 19.0$^1$ & &$5 \%$\\
\object{1RXS J121158.1+224236}    & 12 11 58.7 & +22 42 32 & 0.455 & 2.61 &  20.2 & 19.6$^1$ & &$7 \%$\\
\object{ON 325} / B2 1215+30      & 12 17 52.0 & +30 07 02 & 0.130 &11.79 &
572.7 & 15.6 & 12.0 & \\
\object{PG 1218+304}              & 12 21 21.8 & +30 10 37 & 0.182 & 9.14 &
71.5 & 17.7 & 13.3 &\\
\object{ON 231}$^*$ / W Comae$^*$ & 12 21 31.7 & +28 13 58 & 0.102 & 1.00 & 732.1 & 16.5 & &\\
\object{1RXS J122424.2+243618}$^*$& 12 24 24.2 & +24 36 24 & 0.218 & 0.96 &  25.9 & 17.7 & 13.7&\\
\object{1RXS J123014.2+251805}    & 12 30 14.0 & +25 18 07 & 0.135 & 1.34 &
244.0 & 16.0$^1$ & 11.7 & \\
\object{MS 1229.2+6430}           & 12 31 31.5 & +64 14 16 & 0.164 & 1.97 &
58.8 & 18.0$^1$ & 13.8 & \\
\object{1RXS J123144.7+284754}$^*new$& 12 31 43.9 & +28 47 51 & 0.236 & 0.83 &
141.5 & 17.5 & 14.6 & $-2 \%$\\
\object{1RXS J123705.6+302002}    & 12 37 06.0 & +30 20 05 & 0.700 & 3.23 &   5.6 & 20.0 & &\\
\object{MS 1235.4+6315}           & 12 37 39.1 & +62 58 41 & 0.297 & 1.62 &
12.5 & 18.9$^1$ & 14.7 & \\
\object{1RXS J123923.1+413245}    & 12 39 22.7 & +41 32 52 & 0.16? & 1.17 &   9.1 & 20.3$^1$ & &\\
\object{1RXS J124141.2+344032}    & 12 41 41.4 & +34 40 31 &       & 1.06 &  10.2 & 20.2$^1$ & &\\
\object{1RXS J124312.5+362743}    & 12 43 12.7 & +36 27 45 &       & 4.66 &
147.9 & 16.6$^1$ & 13.4 &\\
\object{1RXS J124818.9+582031}    & 12 48 18.8 & +58 20 29 &       & 1.73 &
245.3 & 14.9$^1$ & 12.0 &\\
\object{1RXS J125301.0+382629}    & 12 53 00.9 & +38 26 26 & 0.372 & 4.38 &   4.8 & 18.9$^1$ & &\\
\object{1ES 1255+244}             & 12 57 31.9 & +24 12 40 & 0.141 & 4.76 &  14.7 & 15.4 & &\\
\object{1RXS J130255.6+505621}    & 13 02 58.0 & +50 56 18 & 0.688 & 2.79 &   2.8 & 19.3$^1$ & &\\
\object{1RXS J132400.2+573918}    & 13 24 00.0 & +57 39 16 & 0.115 & 1.11 &  44.2 & 17.8$^1$ & 13.5&\\
\object{1RXS J132615.0+293332}    & 13 26 15.0 & +29 33 30 & 0.431 & 1.48 &   5.6 & 18.7$^1$ & &\\
\object{1RXS J134029.9+441007}    & 13 40 29.5 & +44 10 07 & 0.548 & 1.91 &  57.2 & 19.3$^1$ & 15.0&$-1 \%$\\
\object{1RXS J134104.8+395942}    & 13 41 04.9 & +39 59 35 & 0.163 & 3.83 &
88.8 & 18.6$^1$ & 14.3 & \\
\object{1RXS J134551.8+425747}$^*$& 13 45 55.3 & +36 50 14 & 0.255 & 0.41 &
144.8 & 20.3 & 14.6 &$34 \%$\\
\object{1RXS J135328.2+560102}    & 13 53 28.0 & +56 00 55 & 0.370 & 1.31 &
14.9 & 19.1$^1$ & 15.5 &\\
\object{1RXS J140450.2+655433}    & 14 04 49.6 & +65 54 30 & 0.364 & 1.07 &  15.4 & 19.4$^1$ & 15.0&\\
\object{1RXS J141031.6+610020}    & 14 10 31.7 & +61 00 10 & 0.384 & 1.37 &
11.4 & 19.9$^1$ & 15.4 & \\
\object{1E 1415+259}              & 14 17 56.6 & +25 43 25 & 0.237 &10.43 &
89.6 & 16.0 & 13.9 & \\
\object{1RXS J141946.4+542324}$^*$& 14 19 46.6 & +54 23 15 & 0.151 & 0.64 & 788.7 & 15.7 & 12.4&\\
\object{1RXS J142239.1+580159}    & 14 22 39.0 & +58 01 55 & 0.638 &10.03 &  13.2 & 17.9$^1$ & &\\
\object{1RXS J142422.9+343407}$^*new$& 14 24 22.7 & +34 33 57 & 0.571?& 0.80 &  10.0 & 18.3 & &\\
\object{1RXS J142700.5+234803}    & 14 27 00.5 & +23 48 03 &       & 2.04 & 430.1 & 16.4 & &\\
\object{1ES 1426+428}             & 14 28 32.6 & +42 40 21 & 0.129 &21.75 &
58.8 & 16.5 & 13.6 &\\
\object{1RXS J143658.7+563934}$^*new$& 14 36 57.8 & +56 39 25 &       & 1.01 &  21.3 & 18.8 & &\\
\object{1RXS J143917.7+393248}    & 14 39 17.5 & +39 32 43 &       & 5.12 &  42.8 & 16.0 & 13.7&\\
\object{MS 1443.5+6349}           & 14 44 34.9 & +63 36 06 & 0.299 & 1.07 &
18.9 &19.7$^1$ & 15.3 & \\
\object{1RXS J144801.0+360833}    & 14 48 01.0 & +36 08 33 &       & 1.55 &
36.2 & 17.2 & 13.7 &\\
\object{1RXS J144932.3+274630}    & 14 49 32.7 & +27 46 22 & 0.228 & 3.56 &
90.7 & 20.0$^1$ & 14.6 & $15 \%$\\
\object{1RXS J145127.5+635421}$^*$& 14 51 26.0 & +63 54 24 & 0.650 & 0.90 &  10.0 & 19.6 & &\\
\object{1RXS J145603.4+504824}    & 14 56 03.7 & +50 48 25 & 0.480 & 8.60 &   4.0 & 18.6$^1$ & &\\
\object{1RXS J145827.3+483250}    & 14 58 28.0 & +48 32 40 & 0.539 & 2.68 &   3.1 & 20.4$^1$ & &\\
\object{1RXS J150101.7+223812}    & 15 01 01.9 & +22 38 06 & 0.235 & 2.06 &  32.4 & 15.5$^1$ & &\\
\object{1RXS J150842.2+270910}    & 15 08 42.7 & +27 09 09 & 0.270 & 2.99 &  39.9 & 18.8$^1$ & &$6 \%$\\
\object{1RXS J151040.8+333515}    & 15 10 42.0 & +33 35 09 & 0.116 & 1.74 &
8.8 & 18.5$^1$ & 13.8 & $37 \%$\\
\object{1ES 1517+656}             & 15 17 47.5 & +65 25 24 & 0.702 & 8.03 &
37.7 & 16.9$^1$ & 13.6 & \\
\object{1RXS J153202.3+301631}    & 15 32 02.2 & +30 16 29 & 0.064 & 2.71 &  54.4 & 15.5 & &\\
\object{1RXS J153324.9+341640}    & 15 33 24.3 & +34 16 40 &       & 1.32 &  30.0 & 17.9 & &\\
\object{1ES 1533+5320}            & 15 35 00.8 & +53 20 35 & 0.890?& 8.03 &  18.2 & 18.9$^1$ & &\\
\object{1RXS J153528.7+392242}$^*$& 15 35 29.1 & +39 22 47 & 0.257 & 0.07 &
19.7 & 19.8 & 14.6 &\\
\object{1ES 1544+820}             & 15 40 15.7 & +81 55 06 &       & 4.02 &
69.9 & 17.1$^1$ & 14.2 &\\
\object{1RXS J155411.8+241415}$^*new$& 15 54 11.9 & +24 14 28 & 0.301 & 0.95 &
12.7 & 20.4$^1$ & & $13 \%$\\
\object{MS 1552.1+2020}           & 15 54 24.1 & +20 11 25 & 0.222 & 2.40 &
79.4 & 18.5$^1$ & & \\
\end{supertabular}
\label{hblsample1}
\end{center}

\begin{figure*}[h]
\begin{minipage}[t]{15cm}
 \begin{minipage}[t]{7cm}
  \begin{flushleft}
        \psfig{figure=2548.f9,width=7cm}
  \end{flushleft}
 \end{minipage}
 \hfill
 \begin{minipage}[t]{7cm}
 \begin{flushright}
         \psfig{figure=2548.f10,width=7cm}
 \end{flushright}
 \end{minipage}
\end{minipage}

\begin{minipage}[t]{15cm}
 \begin{minipage}[t]{7cm}
  \begin{flushleft}
         \psfig{figure=2548.f11,width=7cm}
  \end{flushleft}
 \end{minipage}
 \hfill
 \begin{minipage}[t]{7cm}
 \begin{flushright}
         \psfig{figure=2548.f12,width=7cm}
 \end{flushright}
 \end{minipage}
\end{minipage}

\begin{minipage}[t]{15cm}
 \begin{minipage}[t]{7cm}
  \begin{flushleft}
        \psfig{figure=2548.f13,width=7cm}
  \end{flushleft}
 \end{minipage}
 \hfill
 \begin{minipage}[t]{7cm}
 \begin{flushright}
         \psfig{figure=2548.f14,width=7cm}
 \end{flushright}
 \end{minipage}
\end{minipage}

\begin{minipage}[t]{15cm}
 \begin{minipage}[t]{7cm}
  \begin{flushleft}
        \psfig{figure=2548.f15,width=7cm}
  \end{flushleft}
 \end{minipage}
 \hfill
 \begin{minipage}[t]{7cm}
 \begin{flushright}
         \psfig{figure=2548.f16,width=7cm}
 \end{flushright}
 \end{minipage}
\end{minipage}

\caption[]{\label{fig:spectra1} Optical spectra of new BL Lac objects and of
  BL Lacs with newly determined redshifts.}
\end{figure*}
\addtocounter{figure}{-1}

\begin{figure*}[h]
\begin{minipage}[t]{15cm}
 \begin{minipage}[t]{7cm}
  \begin{flushleft}
        \psfig{figure=2548.f17,width=7cm}
  \end{flushleft}
 \end{minipage}
 \hfill
 \begin{minipage}[t]{7cm}
 \begin{flushright}
         \psfig{figure=2548.f18,width=7cm}
 \end{flushright}
 \end{minipage}
\end{minipage}

\begin{minipage}[t]{15cm}
 \begin{minipage}[t]{7cm}
  \begin{flushleft}
         \psfig{figure=2548.f19,width=7cm}
  \end{flushleft}
 \end{minipage}
 \hfill
 \begin{minipage}[t]{7cm}
 \begin{flushright}
         \psfig{figure=2548.f20,width=7cm}
 \end{flushright}
 \end{minipage}
\end{minipage}

\begin{minipage}[t]{15cm}
 \begin{minipage}[t]{7cm}
  \begin{flushleft}
        \psfig{figure=2548.f21,width=7cm}
  \end{flushleft}
 \end{minipage}
 \hfill
 \begin{minipage}[t]{7cm}
 \begin{flushright}
         \psfig{figure=2548.f22,width=7cm}
 \end{flushright}
 \end{minipage}
\end{minipage}

\begin{minipage}[t]{15cm}
 \begin{minipage}[t]{7cm}
  \begin{flushleft}
        \psfig{figure=2548.f23,width=7cm}
  \end{flushleft}
 \end{minipage}
 \hfill
 \begin{minipage}[t]{7cm}
 \begin{flushright}
         \psfig{figure=2548.f24,width=7cm}
 \end{flushright}
 \end{minipage}
\end{minipage}

\caption[]{Optical spectra of new BL Lac objects and of
  BL Lacs with newly determined redshifts.}
\end{figure*}
\addtocounter{figure}{-1}
\begin{figure*}[h]
\begin{minipage}[t]{15cm}
 \begin{minipage}[t]{7cm}
  \begin{flushleft}
        \psfig{figure=2548.f25,width=7cm}
  \end{flushleft}
 \end{minipage}
 \hfill
 \begin{minipage}[t]{7cm}
 \begin{flushright}
         \psfig{figure=2548.f26,width=7cm}
 \end{flushright}
 \end{minipage}
\end{minipage}

\begin{minipage}[t]{15cm}
 \begin{minipage}[t]{7cm}
  \begin{flushleft}
         \psfig{figure=2548.f27,width=7cm}
  \end{flushleft}
 \end{minipage}
 \hfill
 \begin{minipage}[t]{7cm}
 \begin{flushright}
         \psfig{figure=2548.f28,width=7cm}
 \end{flushright}
 \end{minipage}
\end{minipage}

\begin{minipage}[t]{15cm}
 \begin{minipage}[t]{7cm}
  \begin{flushleft}
        \psfig{figure=2548.f29,width=7cm}
  \end{flushleft}
 \end{minipage}
 \hfill
 \begin{minipage}[t]{7cm}
 \begin{flushright}
         \psfig{figure=2548.f30,width=7cm}
 \end{flushright}
 \end{minipage}
\end{minipage}

\begin{minipage}[t]{15cm}
 \begin{minipage}[t]{7cm}
  \begin{flushleft}
        \psfig{figure=2548.f31,width=7cm}
  \end{flushleft}
 \end{minipage}
 \hfill
 \begin{minipage}[t]{7cm}
 \begin{flushright}
         \psfig{figure=2548.f32,width=7cm}
 \end{flushright}
 \end{minipage}
\end{minipage}

\caption[]{Optical spectra of new BL Lac objects and of
  BL Lacs with newly determined redshifts.}
\end{figure*}
\addtocounter{figure}{-1}

\begin{figure*}[h]
\begin{minipage}[t]{15cm}
 \begin{minipage}[t]{7cm}
  \begin{flushleft}
        \psfig{figure=2548.f33,width=7cm}
  \end{flushleft}
 \end{minipage}
 \hfill
 \begin{minipage}[t]{7cm}
 \begin{flushright}
         \psfig{figure=2548.f34,width=7cm}
 \end{flushright}
 \end{minipage}
\end{minipage}

\begin{minipage}[t]{15cm}
 \begin{minipage}[t]{7cm}
  \begin{flushleft}
         \psfig{figure=2548.f35,width=7cm}
  \end{flushleft}
 \end{minipage}
 \hfill
 \begin{minipage}[t]{7cm}
 \begin{flushright}
   \hfill
 \end{flushright}
 \end{minipage}
\end{minipage}

\caption[]{Optical spectra of new BL Lac objects and of
  BL Lacs with newly determined redshifts.}
\end{figure*}

\end{document}